\documentclass[amsmath,amssymb,prr,aps,showpieces,superscriptaddress,twocolumn, longbibliograph]{revtex4-2}

\usepackage[utf8]{inputenc}
\usepackage[export]{adjustbox}
\usepackage{tikz}
\usepackage{amsmath}
\usepackage{amssymb}
\usepackage{hyperref}
\usepackage{graphicx}
\usepackage{epstopdf}
\usepackage{dcolumn}
\usepackage{soul}
\usepackage{xcolor}
\usepackage{mathrsfs}
\usepackage{bm}
\usepackage{ulem}
\usepackage{soul}
\usepackage{xspace}
\usepackage{verbatim}
\usepackage{color}
\usepackage{xcolor}
\newcommand{\bra}[1]{\left\langle #1 \right|}
\newcommand{\ket}[1]{\left| #1 \right\rangle}

\hypersetup{colorlinks=true,linkcolor=blue,citecolor=blue, filecolor=blue,urlcolor=blue,breaklinks=true}

\definecolor{lime}{HTML}{A6CE39}
\DeclareRobustCommand{\orcidicon}{%
	\begin{tikzpicture}
	\draw[lime, fill=lime] (0,0) 
	circle [radius=0.16] 
	node[white] {{\fontfamily{qag}\selectfont \tiny ID}};
	\draw[white, fill=white] (-0.0625,0.095) 
	circle [radius=0.007];
	\end{tikzpicture}
	\hspace{-2mm}
}

\foreach \x in {A, ..., Z}{%
	\expandafter\xdef\csname orcid\x\endcsname{\noexpand\href{https://orcid.org/\csname orcidauthor\x\endcsname}{\noexpand\orcidicon}}
}

\begin{document} 

\title[]
{Single excitation swap in a modified Jaynes-Cummings-Hubbard lattice}


\author{M. Ahumada \orcidA{}}
\email{maritza.ahumada@usm.cl}
\affiliation{Departamento de F\'isica, Universidad de Santiago de Chile (USACH), Avenida V\'ictor Jara 3493, 9170124, Santiago, Chile}

\author{N. Valderrama-Quinteros\orcidB{}}
\affiliation{Departamento de F\'isica, Universidad de Santiago de Chile (USACH), Avenida V\'ictor Jara 3493, 9170124, Santiago, Chile}

\author{D. Tancara\orcidC{}}
\affiliation{Departamento de F\'isica, Universidad de Santiago de Chile (USACH), Avenida V\'ictor Jara 3493, 9170124, Santiago, Chile}

\author{G. Romero \orcidD{}}
\affiliation{Departamento de F\'isica, CEDENNA, Universidad de Santiago de Chile (USACH), Avenida V\'ictor Jara 3493, 9170124, Santiago, Chile}

\begin{abstract}    

Controlling the transport and nature of quantum excitations in low-dimensional systems is a key requirement for scalable quantum devices, including communication networks and quantum simulators. We propose a one-dimensional hybrid quantum lattice model, in which each lattice unit integrates a single-mode resonator that interacts with a two-level system (TLS), featuring direct coupling between adjacent TLSs. This configuration enables the coherent propagation of excitations with tunable atomic, photonic, or polaritonic character. Beyond conventional single-excitation transport, we demonstrate that appropriate impedance-matching and resonance conditions allow for the controlled swapping of excitation type as the excitation propagates along the lattice. We analyze the resulting dynamics using local observables and pairwise concurrence to track both transport and quantum correlations. Our results establish a minimal platform for controlled single-excitation conversion, with direct relevance to hybrid quantum networks, on-chip quantum interconnects, and engineered quantum simulators.
\end{abstract}

\maketitle
\section{Introduction}

The ability to control the propagation of quantum excitations in engineered many-body systems is a central requirement for quantum communication networks \cite{Bose2003,Kimble2008,Wehner2018Oct}, quantum simulators \cite{Houck2012,Georgescu2014}, quantum state transfer protocols \cite{Cirac1997Apr,Christandl2004May,Shi2005Mar,Burgarth2005May,Burgarth2005May1,Perez-Leija2013Jan,Ghosh2014Dec,Ashhab2015Dec,Cheng2022Jan,Penas2023Sep,Penas2024Sep}, distributed quantum computing architectures \cite{Cirac1999Jun,Serafini2006Jan,Yin2007Jan,Axline2018Jul,Kurpiers2018Jun,Leung2019Feb,Penas2022May}, and energy transfer devices \cite{Feist2015,Chng2025}. 
One-dimensional (1D) quantum lattices provide an especially attractive setting for these tasks, as they enable controlled transport of single or few excitations while remaining amenable to both analytical methods and numerical simulations. In such lattices, excitations may take different physical forms, including spin waves \cite{Lieb1972,Calabrese2012}, bosons \cite{Fisher1989,Cheneau2012}, hybrid light-matter excitations \cite{Greentree2006Dec,Hartmann2006Dec,Angelakis2007Sep,Makin2009Oct}, and photons \cite{Shen2005,Chang2007,Roy2017}. Nowadays, 1D quantum lattices can be implemented in circuit quantum electrodynamics \cite{Li2018Nov,Xiang2024Jun,Roy2025Mar}, trapped ions \cite{Zhang2017,Ohira2021,Muralidharan2023}, and Rydberg atoms \cite{Bernien2017}, enabling the realization of the hybrid architectures in which light-matter coupling and inter-qubit interactions can be engineered with high precision.

A paradigmatic framework capturing the dynamics of the photonic, atomic, or hybrid excitations is provided by the Jaynes-Cummings-Hubbard (JCH) model \cite{Greentree2006Dec,Hartmann2006Dec,Angelakis2007Sep,Ogden2008,Chakrabarti_2011}. Originally introduced in the context of quantum phase transitions, the JCH model has since been explored in a wide range of static and dynamical regimes \cite{Koch2009Aug,Makin2009Oct,Nietner2012Apr,Hayward2012Jun,Bujnowski2014Oct,Xue2017Nov,Figueroa2018Aug,Prasad2018Nov,Pena2020Mar,Li2021Jul,Norambuena2022,Ma2022Apr,Li2022Sep,Tudorovskaya2024Mar,PRXQuantum.5.010339}. It has also been proposed as a platform for quantum state transfer in both long-range \cite{Almeida2016quantum} and short-range lattice geometries \cite{Baum2022May,Yue2024Nov}. These works have established the feasibility of coherent single-excitation transport in hybrid light-matter systems. However, the ability to dynamically swap the character of an excitation during its propagation, for example, an atomic excitation into a polaritonic one and vice versa, remains unexplored, despite its potential relevance for hybrid quantum networks and on-chip quantum interconnects.

In this work, we address this gap by proposing a one-dimensional hybrid lattice in which each unit consists of a single-mode resonator locally coupled to a two-level system, while neighbouring units interact directly through their TLSs, see the schematic of Fig.\ref{Fig1}. This architecture is a modified JCH model that naturally supports the propagation of polaritonic and purely atomic excitations. However, purely photonic excitations require adding an ancillary emitter that mediates the conversion between excitation types. We prove that, under suitable impedance-matching and resonance conditions derived in the hybrid light-matter basis, the system enables a controlled swapping of excitation type at a local interface within the lattice. To establish this mechanism, we revisit how a single atomic excitation, initially prepared in the emitter, can be injected into the lattice and converted into propagating purely atomic, purely photonic, or polaritonic excitations by parameter tuning. This analysis provides the necessary framework to identify the conditions under which coherent excitation-type swapping occurs dynamically during propagation. 
The dynamical swapping and conversion mechanisms are studied by using the average of local observables and entanglement via pairwise concurrences \cite{Wootters1998}. 

The proposed mechanism relies on parameter regimes that are accessible in circuit-QED architectures \cite{circuitQED}, making it experimentally feasible with current technology. Our results establish a minimal and versatile platform for controlled single-excitation conversion and swapping, with applications to the design of controllable quantum links \cite{Penas2023Sep,Penas2024Sep,barahona2025}, hybrid quantum networks, and engineered quantum simulators of light-matter systems.

\section{Model}

\begin{figure}
\centering
\includegraphics[width=\columnwidth]{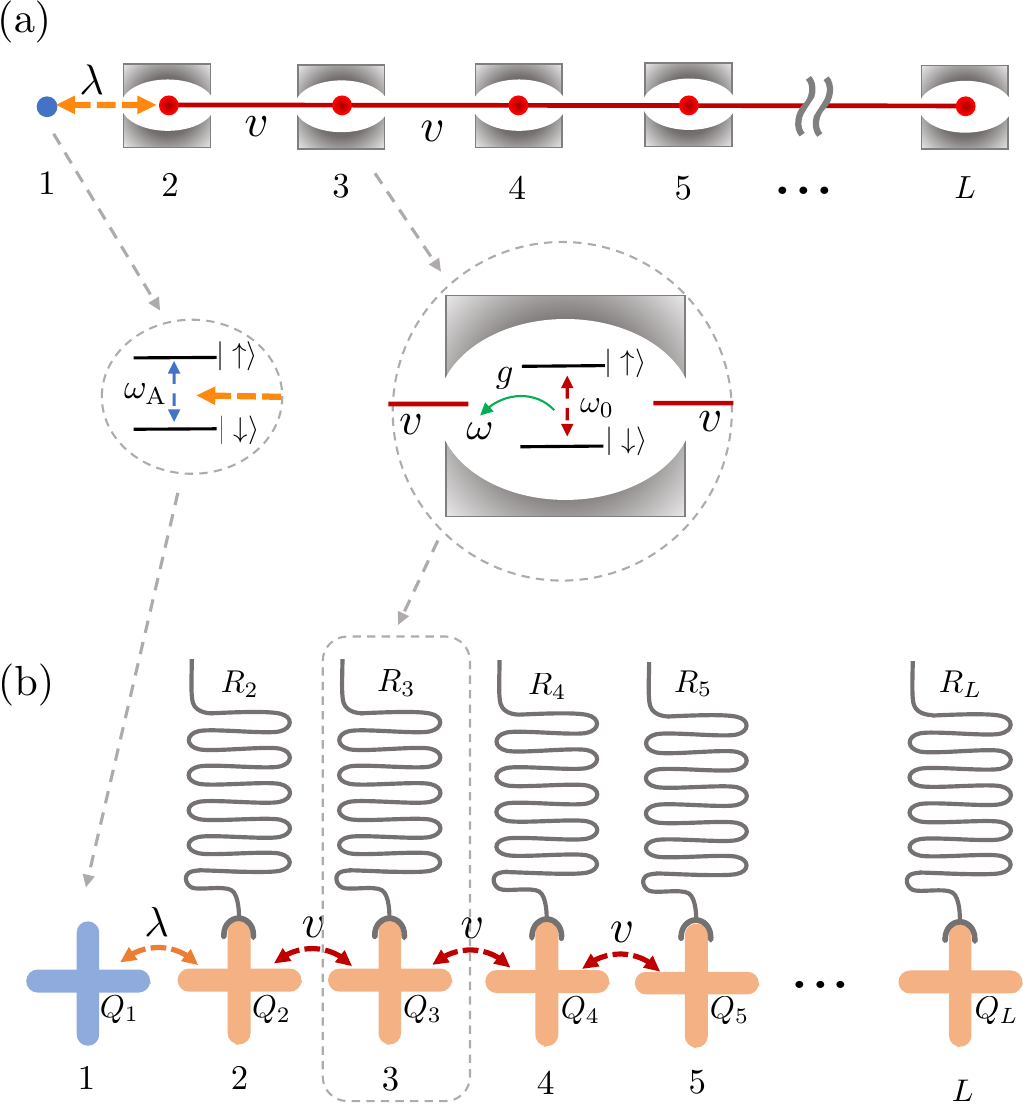}
\caption{(a) Schematic representation of hybrid quantum lattice. The blue dot represents the activation qubit (left inset) with frequency $\omega_{\rm{A}}$, which is coupled to the TLS of the first unit of the hybrid lattice with coupling strength $\lambda$ (orange segmented arrow). The right inset shows a single-mode resonator interacting with a TLS. The TLSs interact with their nearest neighbors with coupling strength $v$ (red solid line). (b) Schematic of a possible physical realization of our model in a circuit QED platform. The latter consists of an array of capacitively coupled transmons represented by the light-blue cross and the light-orange crosses. Each transmon is capacitively coupled to a resonator depicted as wavy lines.}
\label{Fig1}
\end{figure}

The hybrid quantum lattice consists of light-matter units, each comprising a single-mode resonator interacting with a single two-level system, as shown schematically in Fig.~\ref{Fig1}(a). In addition to the light-matter interaction, the TLSs interact directly with their nearest neighbors. An \textit{activation qubit} (emitter) is placed at the first site of the hybrid quantum lattice, and its primary role is to inject a single excitation into the hybrid lattice. The activation qubit is directly coupled to the TLS of the first unit in the hybrid lattice, located at site $j=2$.
The hybrid quantum lattice supports the propagation of three types of single excitations: polaritons, purely atomic excitations, and purely photonic excitations. The proposed model is inspired by state-of-the-art circuit QED experiments  \cite{Li2018Nov,Zha2020,Zhao2022,Song2024} as schematically shown in Fig.~\ref{Fig1}(b). Here, a set of transmons, represented by crosses, are capacitively coupled in a linear array. Each light-orange transmon is capacitively coupled to a single-mode resonator (wavy lines). We stress that in Refs.\cite{Li2018Nov,Zha2020,Zhao2022,Song2024}, the resonators were used for measuring the transmon state. However, there are no fundamental physical limitations preventing the transmon-resonator system from strongly interacting and forming polaritons.      

The system Hamiltonian consists of three contributions, $H=H_{\rm A}+H_{\rm JC}+H_{\rm I}$. Here, $H_{\rm A}$ describes the activation qubit and its interaction with the hybrid lattice composed of light-matter units. $H_{\rm JC}$ denotes the interaction between each single-mode resonator and its corresponding TLS via the Jaynes-Cummings model \cite{Jaynes1963,Larson2024Aug,Larson2024}. $H_{\rm I}$ accounts for the nearest-neighbor interaction between the TLSs. These terms are given by

\begin{equation}
    H_{\rm A}= \hbar \omega_{\rm{A}}\,\hat{\sigma}^{+}_1\hat{\sigma}^{-}_1+\hbar\lambda\left(\hat{\sigma}_{1}^{+}\hat{\sigma}_{2}^{-}+\hat{\sigma}_{1}^{-}\hat{\sigma}_{2}^{+}\right)\,,
\end{equation}
\begin{equation}
    H_{\rm JC}=\sum_{j=2}^{L} \hbar\left[\omega \,\hat{a}^{\dagger}_{j}\hat{a}_{j}+\omega_{0}\,\hat{\sigma}^{+}_{j}\hat{\sigma}^{-}_{j}+g\,(\hat{\sigma}^{+}_{j}\hat{a}_{j}+\hat{\sigma}^{-}_{j}\hat{a}^{\dagger}_{j})\right]\,,
\end{equation}
\begin{equation}
    H_{\rm I}=\sum_{j=2}^{L-1} \hbar v\left(\hat{\sigma}_{j}^{+}\hat{\sigma}_{j+1}^{-}+\hat{\sigma}_{j}^{-}\hat{\sigma}_{j+1}^{+}\right)\,,
    \label{HII}
\end{equation}
where $\hat{a}_{j}^{\dag}$ ($\hat{a}_{j}$) is the creation (annihilation) operator of a single bosonic excitation in the $j$-th resonator with frequency $\omega$, and 
$\hat{\sigma}_{j}^{+}$ and $\hat{\sigma}^{-}_{j}$ are the raising and lowering operators of the $j$-th TLS in the eigenbasis $\{|\uparrow_{j}\rangle,|\downarrow_{j}\rangle\}$. Additionally, $\omega_{\rm{A}}$ denotes the frequency of the activation qubit,  $\omega_0$ the transition frequency of the TLSs in the hybrid lattice, $g$ the light-matter coupling strength, $\lambda$ the coupling strength between the activation qubit and the TLS in the first unit of the hybrid lattice, and $v$ the coupling strength between nearest-neighbor TLSs. Along this work, we deal with a single excitation propagation. In this case, the Hamiltonian of our model can be mapped to an effective single-particle problem and is therefore formally related to the JCH systems studied in Ref.~\cite{Makin2009Oct}.

To determine impedance-matching conditions and analyze the propagation of different types of single excitations, it is convenient to rewrite the full Hamiltonian in the basis of hybrid light-matter states. The methodology for performing the basis transformation has been reported previously for the Jaynes-Cummings-Hubbard model in Refs.~\cite{Angelakis2007Sep,Koch2009Aug}. Then, we denote the upper $(+)$ and lower $(-)$ polaritonic basis states are
\begin{equation}
    |n,\alpha\rangle_j = \gamma_{n\alpha} |\downarrow,n\rangle_j+\rho_{n\alpha}|\uparrow,n-1\rangle_j \label{pol_states}\,,
\end{equation}
where $\alpha = \pm$ denotes the polaritonic branches. The coefficients are defined by 
\begin{eqnarray}\label{coefficients}
    &&\rho_{n+}=\cos\bigg(\frac{\theta_n}{2}\bigg),\quad \gamma_{n+}=\sin\bigg(\frac{\theta_n}{2}\bigg),\nonumber\\
    &&\rho_{n-}=-\gamma_{n+},\quad\rho_{n+}=\gamma_{n-}\,,
\end{eqnarray}
and the mixing angle $\theta_{n}$ is given by $\tan(\theta_n)=2g\sqrt n/\Delta$ with $\Delta = \omega_0-\omega$ as the detuning.

The energy eigenvalues associated with these eigenstates are
\begin{equation}
    E^{\pm}_n/\hbar = n\omega+\frac{\Delta}{2}\pm\frac{\Delta}{2}\sqrt{1+\frac{4ng^2}{\Delta^2}}\,.
\label{Eigenenergies}
\end{equation}

The polaritonic creation operator at the $j$th site is $P^{\dagger(n,\alpha)}_j = |n,\alpha\rangle_j\langle0,-|$, where  $|\downarrow,0\rangle = |0,-\rangle$ is the ground state and the unphysical state $|0,+\rangle = |\textbf{\O}\rangle$ is a null vector. The previous determines the values of $\gamma_{0-}=1$ and $\gamma_{0+}=\rho_{0\alpha}=0$.

The terms of the Hamiltonian written in the polaritonic basis are

\begin{equation}
    \begin{split}
        H_{\rm A} =\hspace{0.1cm}&\hbar\omega_{\rm{A}}\,\hat{\sigma}^{+}_1\hat{\sigma}^{-}_1+\\&\hbar\lambda \hat{\sigma}_{1}^{+}\sum_{n=0}^{\infty}\sum_{\alpha,\beta=\pm}\rho_{n\alpha}\gamma_{n-1\beta}\hat{P}_{2}^{\dagger(n-1,\beta)}\hat{P}_{2}^{(n,\alpha)}+\\&\hbar\lambda \hat{\sigma}_{1}^{-}\sum_{n=0}^{\infty}\sum_{\alpha,\beta=\pm}\rho_{n\alpha}\gamma_{n-1\beta}\hat{P}_2^{\dagger(n,\alpha)}\hat{P}_2^{(n-1,\beta)},\hspace{0.1cm}
    \end{split}
    \label{HA}
\end{equation}
\begin{equation}
    \begin{split}
H_{\rm JC}=\hspace{0.1cm}\hbar\sum_{j=2}^L\sum_{n=0}^\infty\sum_{\alpha=\pm}E_n^\alpha \hat{P}_j^{\dagger(n,\alpha)}\hat{P}_j^{(n,\alpha)},
    \end{split}
    \label{HJC}
\end{equation}
\begin{equation}
    \begin{split}
    H_{\rm I} =\hspace{0.1cm}&\hbar v\sum_{j=2}^{L-1}\sum_{n,m=1}^{\infty} \sum_{\substack{\alpha,\alpha'=\pm\\\beta,\beta'=\pm}}\big[ \, t_{n}^{\alpha\beta}t_{m}^{\alpha'\beta'}\hat{P}_{j}^{\dagger(n,\alpha)}\hat{P}_{j}^{(n-1,\beta)}\\
&\hat{P}_{j+1}^{\dagger(m-1,\beta')}\hat{P}_{j+1}^{(m,\alpha')}+ {\rm h.c.} \, \big],
    \end{split}
        \label{HI}
\end{equation}

where $t_{n}^{\alpha\beta}=\rho_{n\alpha}\gamma_{n-1\beta}$. See Appendix~\ref{ApendixA} for details.

\section{Single-excitation propagation in atomic, photonic, and polaritonic channels}

Previous studies have examined the three propagation regimes of the JCH model in different contexts. The atomic propagation regime was analysed by Ogden et al. in Ref.~\cite{Ogden2008}, using a minimal two-site model and a description in terms of delocalized field modes, where atomic excitation transfer can occur with negligible field population. A generalization of this mechanism to larger arrays was later presented by Chakrabarti et al. in Ref.~\cite{Chakrabarti_2011}, who used delocalized photonic and atomic modes to describe predominantly atomic or photonic propagation under appropriate detuning conditions. Photonic and hybrid excitation transport has also been studied in Ref.~\cite{Makin2009Oct}, where the atomic signal corresponds to the atomic component of a hybrid excitation. In addition, Ref.~\cite{Almeida2016quantum} investigated atomic-state transfer mediated indirectly through hybrid excitation in a staggered JCH lattice, where the excitation acquires a non-negligible photonic component in the bulk during its propagation.

In this section, we present the propagation of different types of excitations, that is, polaritons, atomic waves, and photons in our model. This analysis provides a framework for proposing the main contribution of our work, namely, the excitation-swap mechanism.
Our methodology is not focused on delocalized collective fields, but rather derives an impedance-matching and resonance condition formulated using the Hamiltonian transformation to the polaritonic basis, along with a large detuning of the local fields. 
In our model, the propagation dynamics of each excitation type strongly depend on suitable impedance-matching and resonance conditions. As will be shown later, by satisfying these conditions and suppressing the excitation propagation through the upper polaritonic branch, we can select the type of information carrier propagating in the system.
We focus on two distinct cases of light-matter interaction: the resonant and dispersive regimes. In the resonant regime, where $\Delta=\omega_{0}-\omega=0$, polaritons dominate the dynamics. In contrast, in the dispersive regime, where $|\Delta|\gg g$, the single photon and atomic-wave excitation become the dominant information carriers. 

We perform numerical simulations using the TEBD algorithm \cite{Vidal2004,Daley2004,Schollwock2011}.
We approximate the evolution operator using the Suzuki-Trotter decomposition \cite{hatano2005}
\begin{equation}
    e^{-i\tau(F+G)}\approx \prod_{l=1}^{k}\,e^{-i\tau(c_{l}F)}\,e^{-i\tau(d_{l}G)}+O(\tau^{n+1}),
\label{Approximant}    
\end{equation}
where $F$ and $G$ contain the sum of the Hamiltonian terms for the even and odd bonds \cite{Vidal2004}, respectively. In particular,  we use the second-order ($n=2$) approximant, where the expansion parameters of Eq.~(\ref{Approximant}) are $k=2$, $c_{1}=1/2$, $d_{1}=1$, $c_{2}=1/2$, $d_{2}=0$. Besides, we consider a time-step $\tau=10^{-3}v^{-1}$ for atomic and polariton propagation, and  $\tau=5\times10^{-3}v^{-1}$ for photon propagation, the total system size of $L=26$ sites, the accessible number states in each resonator is $n_{\rm max}=2$, and the maximum bond dimension is $\chi_{\rm max}=4$. Within the simulated time, the latter ensures a truncation error $\epsilon_{\rm trunc}\lesssim 10^{-8}$, see Appendix \ref{AppendixB}. We emphasize that the system evolves within the single-excitation subspace, allowing for an equivalent description in terms of a finite-dimensional Hamiltonian. Nevertheless, we employ the TEBD algorithm as a flexible and scalable numerical approach that can readily be extended to higher numbers of excitations, involving larger local Hilbert spaces. In this sense, TEBD provides a consistent framework for future studies of multi-excitation dynamics.

We use an initial condition that enables the activation of the hybrid quantum lattice. The activation qubit can undergo a transition from $|\uparrow_{1}\,\rangle$ to $|\downarrow_{1}\,\rangle$, thereby transferring a single excitation. This initial state is given by
\begin{equation}
|\Psi_{0}\rangle=|\uparrow_{1}\,\rangle\otimes|\downarrow,\,0\,\rangle_{2}\otimes|\downarrow,\,0\,\rangle_{3}\otimes\dots\otimes|\downarrow,\,0\,\rangle_{L}\,.
\label{IniCond2}
\end{equation}

In the following, we present the dynamics of each type of single excitation and give details on the conditions under which they emerge.

\subsection{Resonant regime: polariton propagation}\label{Sec_Polariton}
\begin{figure}[t]
\centering
\includegraphics[width=\columnwidth]{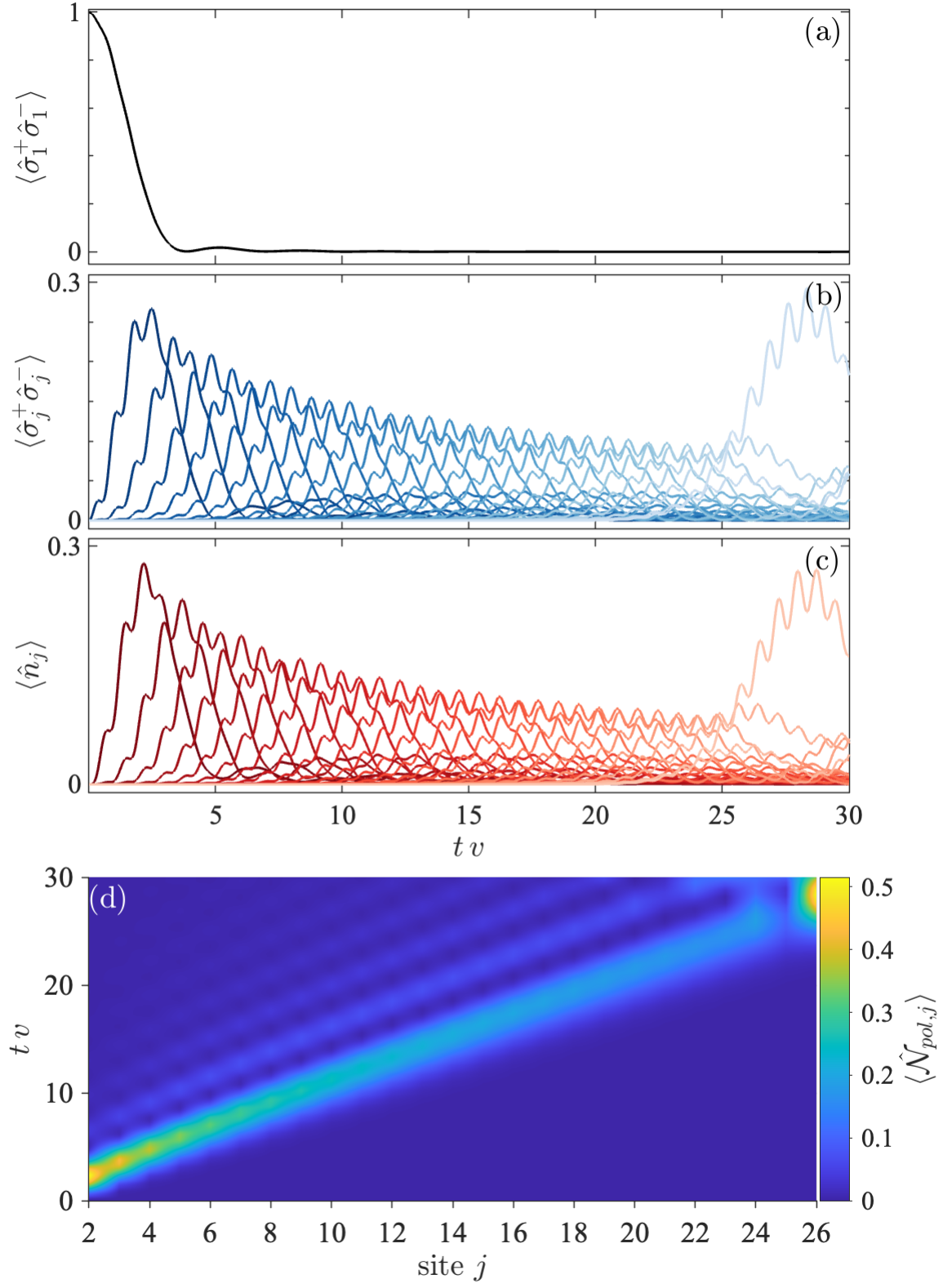}
\caption{Polariton-wave propagation: (a) Time evolution of the average excitation number in the activation qubit. (b) Time evolution of the average excitation number in the $j$th TLS of the hybrid lattice, for $2\leq j \leq 26$. (c) Average photon number as a function of time. (d) Average polariton number as a function of time and the site index $j$. The parameters are $\Delta=0$ with $\omega=\omega_{0}=1$, $v=0.01\,\omega$, $g=4\,v$, $\omega_{\rm{A}}=\omega_{pol}=\omega-g$, $\lambda=-v/\sqrt{2}$, $\tau=10^{-3}v^{-1}$, $\chi_{\rm max}=4$, and $L=26$.}
\label{Fig2}
\end{figure}

Hereafter, and for the sake of simplicity, we will consider the lower polaritonic branch for a single polariton propagating along the network. The latter is possible if three conditions are fulfilled. 
\begin{enumerate}
\item  The propagation through the upper polaritonic branch is suppressed if $g\gg \lvert v\lvert/4$. The derivation of this condition is provided in Appendix~\ref{ApendixA}. Appendix~\ref{AppendixC} provides a numerical analysis confirming the suppression of propagation in the upper polaritonic branch under this condition.

\item The frequency of the activation qubit must be resonant with the transition frequency of the lower polaritonic branch, i.e., $\omega_{\rm{A}}=\omega-g$. From Eq.~(\ref{Eigenenergies}), the energy levels in the lower polaritonic branch are given by
\begin{equation}
    E^{-}_n/\hbar = n\omega+\frac{\Delta}{2}-\frac{1}{2}\sqrt{\Delta^2+4ng^2}\,.
\label{EnergyBranch-}    
\end{equation}
In the resonant regime, $\Delta=0$, and considering a single excitation, $n=1$, Eq.~(\ref{EnergyBranch-}) reduces to
\begin{equation}
    E^{-}_{1}/\hbar = \omega-g = \omega_{pol}\,.
\end{equation}

\item The effective coupling between the activation qubit and the first polariton must match the effective coupling between two nearest-neighbor polaritons. This conditions leads to $\lambda=-v/\sqrt{2}$.

\end{enumerate}

To determine the third condition, we focus on the Hamiltonian $H_{\rm I}$ in the representation of the polaritonic basis, Eq.~(\ref{HI}), which accounts for the polariton hopping between nearest-neighbour sites. In the case of a single excitation propagating through the lower polaritonic branch, the indexes correspond to $n=1$, $\alpha=\beta=-$, $m=1$, $\alpha^{\prime}=\beta^{\prime}=-$, and the effective coupling between polariton nearest-neighbour reads
\begin{equation}
\tilde{v} = v\,t_{1}^{--}\,t_{1}^{--}=v\,(\rho_{1-}\,\gamma_{0-})^{2}=v\,(\rho_{1-})^{2}\,.
\label{Acop_vest_polariton}
\end{equation}
Let us focus on the Hamiltonian $H_{\rm A}$ in Eq.~(\ref{HA}). From this expression, the effective coupling between the activation qubit and the first polariton is given by
\begin{equation}
\tilde{\lambda} = \lambda\,\rho_{1-}\,\gamma_{0-}=\lambda\,\rho_{1-}\,,
\end{equation}
Both effective coupling matches, resulting
\begin{equation}
\tilde{\lambda}=\tilde{v}\,\Rightarrow\,\lambda=v\,\rho_{1-}\,,
\label{condition3}
\end{equation}
where using Eqs.~(\ref{coefficients})
\begin{equation}
    \rho_{1-}=-\sin{\left(\frac{\theta_{1}}{2}\right)},\hspace{0.2cm} \text{with} \hspace{0.2cm} \tan{(\theta_{1})}=\frac{2g}{\Delta}\,.
\label{coefficient_condition3}
\end{equation}

The match impedance condition Eqs.~(\ref{condition3}) is valid for any $\Delta$. In the particular case of the resonant regime, $\Delta=0$ and $\tan{(\theta_{1})}$ diverge at $\theta_1=\pi/2$, resulting in $\rho_{1-}=-1/\sqrt{2}$, and the condition reduces to $\lambda=-v/\sqrt{2}$.

Tuning the lattice parameters to satisfy the three previously described conditions enables the propagation of a single polariton through the hybrid quantum lattice. Figure~\ref{Fig2} shows the mechanism of single-excitation generation, its release, and subsequent polariton propagation. Figure~\ref{Fig2}(a) displays the behavior of the activation qubit, where the average excitation number $\langle\hat{\sigma}^{+}_{1}\hat{\sigma}^{-}_{1}\rangle=1$ at $t=0$, and eventually decays to 0, injecting a single excitation into the hybrid system. Once released, the excitation propagates to the right as a single polariton. Figures~\ref{Fig2}(b) and \ref{Fig2}(c) show the average excitation number of the $j$th TLS and the average photon number in each resonator, respectively. Since the polariton number operator is defined as $\hat{\mathcal{N}}_{pol,j}= \hat{\sigma}^{+}_{j}\hat{\sigma}^{-}_{j}+  \hat{n}_{j}$, these two figures separately depict the respective contributions of atomic and photon components to the polariton dynamics. In Fig.~\ref{Fig2}(d), we show the average polariton number $\langle\hat{\mathcal{N}}_{pol,j} \rangle$ as a function of time and site index $j$. The polaritonic excitation appears as a dispersive pulse, with a decreasing amplitude and increasing width over time.

\subsection{Dispersive regime: photon propagation}

\begin{figure}[t]
\centering
\includegraphics[width=\columnwidth]{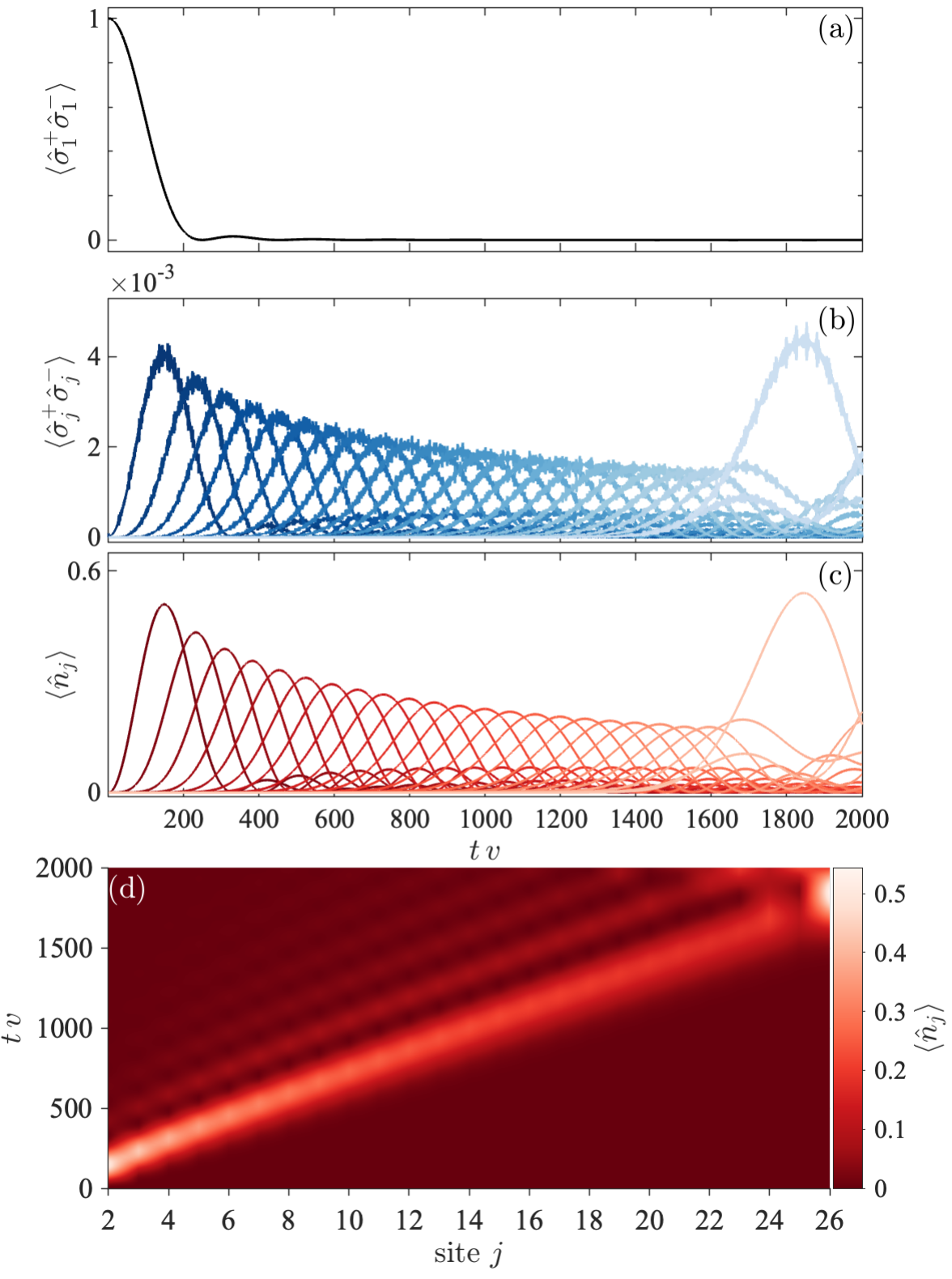}
\caption{Photon-wave propagation: (a) Time evolution of the average excitation number in the activation qubit. (b) Time evolution of the average excitation number in the $j$th TLS of the hybrid lattice, for $2\leq j \leq 26$. (c) Average photon number as a function of time. (d) Average photon number as a function of time and the site index $j$. The parameters are $\omega=10$, $\omega_{0}=1$, $\Delta=\omega_{0}-\omega$, $g=0.08\,\omega$, $\omega_{\rm A}=\omega-g^2/\Delta$, $v=0.05\,\omega$, $\lambda=-v\,\sin{\left(\theta_{1}/2\right)}$ with $\theta_{1}=\arctan{(2g/\Delta)}$, $\tau=5\times10^{-3}v^{-1}$, $\chi_{\rm max}=4$, and $L=26$.}
\label{Fig3}
\end{figure}
%

In the dispersive regime ($|\Delta|\gg g$), purely photonic transport may be mediated by the hybrid light-matter excitations. For photon propagation to occur, the following conditions must be satisfied:

\begin{enumerate}
\item The condition $g\gg \lvert v/4\lvert$ ensures the suppression of propagation through the upper polaritonic branch. 

\item The activation qubit frequency must be resonant with the transition frequency of the lower polaritonic branch. In the dispersive regime, the condition is now $\omega_{\rm A}=\omega-g^2/\Delta$. From Eq.~(\ref{EnergyBranch-}), the energy levels in the lower polaritonic branch reads
\begin{equation}
    E^{-}_n/\hbar = n\omega+\frac{\Delta}{2}-\frac{\Delta}{2}\sqrt{1+\chi(n)}\,,\quad\chi(n)=\frac{4g^2n}{\Delta^{2}}.
\label{Energybranch-_exp}    
\end{equation}
In the dispersive regime, $|\Delta|\gg g$ and the quantity $\chi(n)\ll1$, allowing to expand the square root in the third term of Eq.~(\ref{Energybranch-_exp}) as a Taylor series around $\chi(n)$. This yields,
\begin{equation}
    E^{-}_n/\hbar \simeq n\omega+\frac{\Delta}{2}-\frac{\Delta}{2}\left(1+\frac{\chi(n)}{2}\right)\, = \left(\omega-\frac{g^2}{\Delta}\right)n\,.
\label{Eigenenergies2}
\end{equation}
Thus, considering a single excitation $n=1$, the energy Eq.~(\ref{Eigenenergies2}) reduces to
\begin{equation}
E^{-}_1/\hbar=\omega-\frac{g^2}{\Delta}
\end{equation}
\item The effective coupling between the activation qubit and the first polariton must match the effective coupling between two nearest-neighbor polaritons, $\lambda=v\,\rho_{1,-}$ [see Eqs.~(\ref{condition3}) and (\ref{coefficient_condition3})].
\end{enumerate}

Single-photon propagation in the hybrid system is illustrated in Fig.~\ref{Fig3}. By tuning the system parameters to satisfy the last three conditions associated with this case, photon-like excitations dominate the transport dynamics. As in the polariton case, the excitation originates from the activation qubit and is released into the system. Once injected, it propagates to the right, now as a single photon, through the hybrid lattice, even though the resonators are not directly coupled to each other.
Figures~\ref{Fig3}(b) and \ref{Fig3}(c) show the average excitation number of the $j$th TLS and the average photon number in each resonator, respectively. In this case, the values of $\langle\hat{\sigma}^{+}_{j}\hat{\sigma}^{-}_{j}\rangle$ in Fig.~\ref{Fig3}(b) remain close to zero, while in Fig.~\ref{Fig3}(c) the photon number reveals a propagating excitation pulse through the resonators. Figure~\ref{Fig3}(d) provides a broader view of the photon propagation across the hybrid lattice.

\begin{figure}[t]
\centering
\includegraphics[width=\columnwidth]{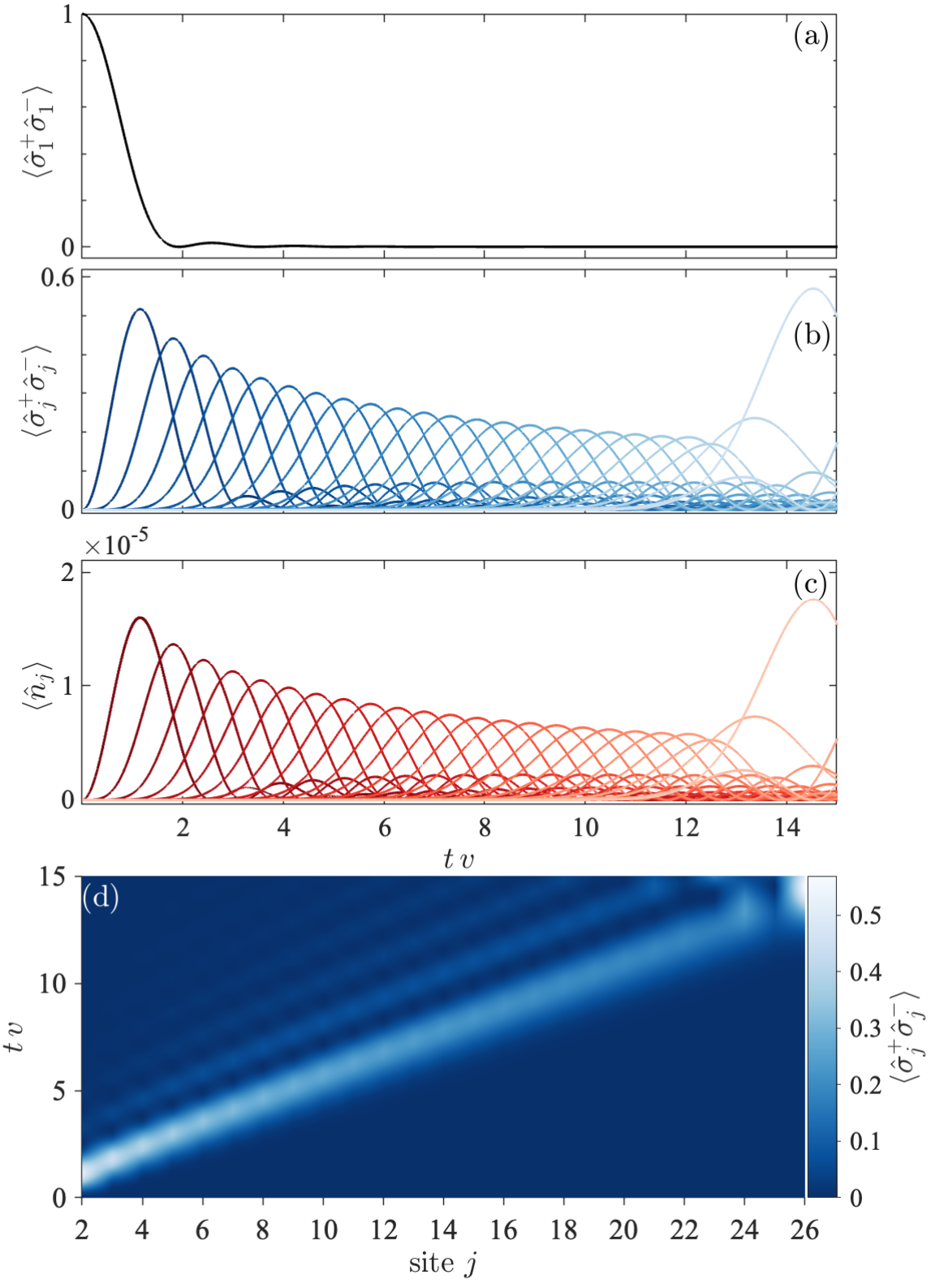}
\caption{Atomic-wave propagation: (a) Time evolution of the average excitation number in the activation qubit. (b) Time evolution of the average excitation number in the $j$th TLS of the hybrid lattice, for $2\leq j \leq 26$. (c) Average photon number as a function of time. (d) Average excitation number in the $j$th TLS as a function of time and the site index $j$. The parameters are $\omega=1$, $\omega_{0}=10$, $\Delta=\omega_{0}-\omega$, $\omega_{\rm{A}}=\omega_{0}$, $g=0.05\,\omega$, $v=0.02\,\omega$, $\lambda=v$, $\tau=10^{-3}v^{-1}$, $\chi_{\rm max}=4$, and $L=26$.}
\label{Fig4}
\end{figure}

\subsection{Dispersive regime: Atomic wave propagation}

In the dispersive regime ($|\Delta|\gg g$), a second case of interest is the propagation of atomic-wave excitations. This regime allows for excitation transport through the TLS sublattice, while the photonic component remains largely unexcited. The following conditions must be satisfied:

\begin{enumerate}
\item The condition $g\gg \lvert v/4\lvert$ ensures the suppression of propagation through the upper polaritonic branch.
\item The frequency of the activation qubit must be resonant with the transition frequency of the TLSs in the hybrid lattice, i.e., $\omega_{\rm{A}}=\omega_{0}$.
\item The coupling between the activation qubit and the TLS in the first resonator unit must match the coupling strength between neighboring TLSs, i.e., $\lambda=v$.
\end{enumerate}

Atomic-wave propagation under these conditions is illustrated in Fig.~\ref{Fig4}. Once released, the excitation propagates through the lattice via the TLS sublattice, forming an atomic-wave [see Fig.~\ref{Fig4}(b)]. In this case, the average photon number in each resonator remains negligible across all sites [see Fig.~\ref{Fig4}(c)]. Figure~\ref{Fig4}(d) provides a global view of the excitation dynamics, showing the transport of the atomic-wave through the hybrid lattice.

\section{Excitation Swap}

\begin{figure}
\centering
\includegraphics[width=\columnwidth]{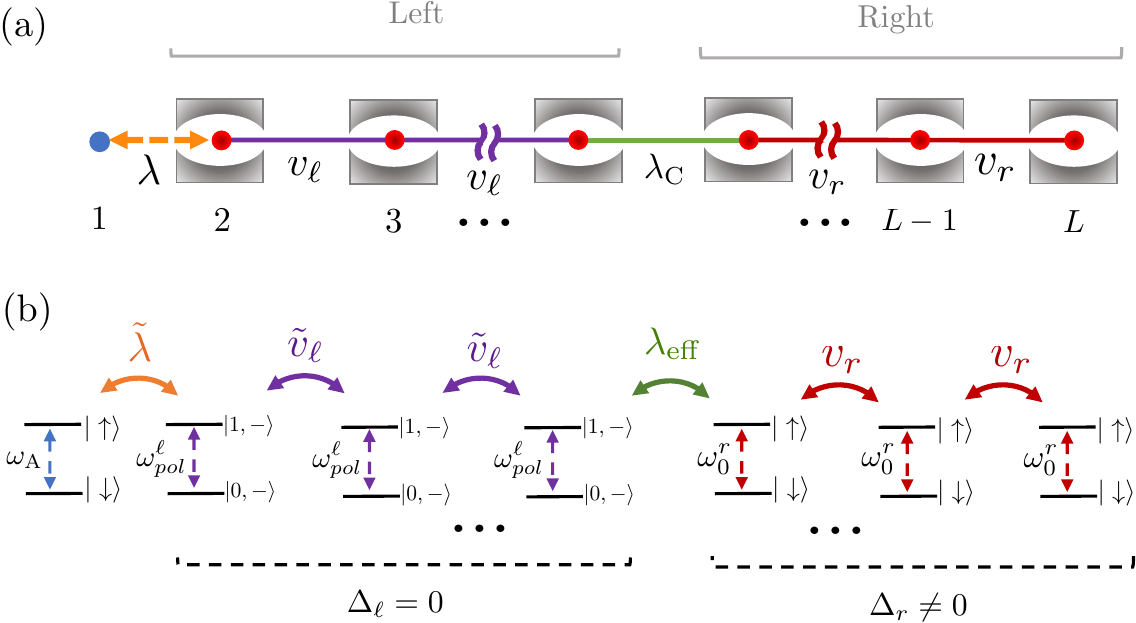}
\caption{(a) Schematic of the hybrid quantum lattice divided into two sections (left and right), each characterized by distinct parameter sets. (b) Schematic representation of the effective energy levels, effective coupling, and effective impedance-matching interface (green curve arrow), enabling an excitation swap of the type polariton $\rightarrow$ atomic-wave can occur. The polariton transition frequencies and effective couplings between nearest-neighbour polaritons are indicated with purple arrows, while those corresponding to the atomic-wave are shown in red. The orange arrow indicates the effective coupling between the activation qubit and the first polariton.}
\label{Fig5}
\end{figure}

\begin{figure*}[t]
\centering
\includegraphics[width=\textwidth]{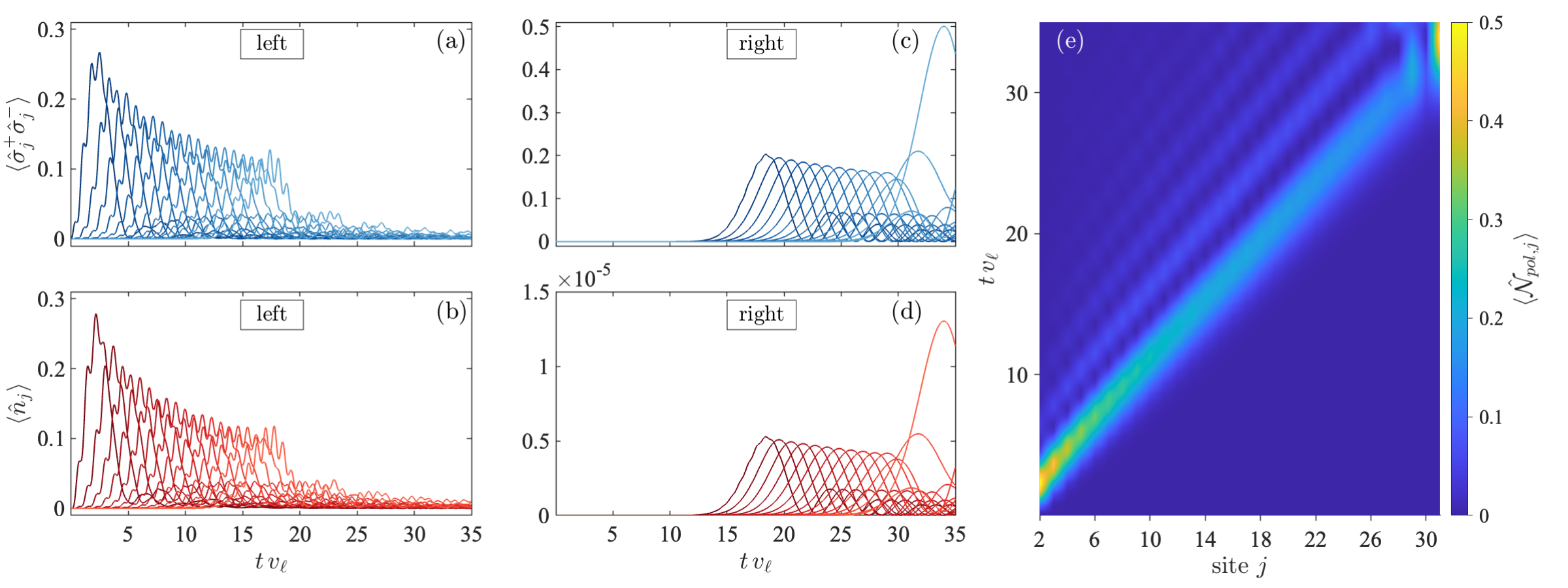}
\caption{Dynamic excitation swap, polariton $\rightarrow$ atomic-wave: Average excitation number of the $j$th TLS and the average photon number as functions of time for the left section ($\ell$), shown in panels (a) and (b), and for the right section ($r$), shown in panels (c) and (d), respectively. (e) Average polariton number as a function of time and the site index $j$ (with $2\leq j \leq 31$).
The parameters in the left section are $\omega_{\ell}=\omega^{\ell}_{0}=1$, $\Delta_{\ell}=\omega^{\ell}_{0}-\omega_{\ell}=0$, $v_{\ell}=0.05\,\omega_{\ell}$, $g_{\ell}=4v_{\ell}$, $\omega_{\rm{A}}=\omega^{\ell}_{pol}=\omega_{\ell}-g_{\ell}$, $\lambda=\lambda_{\rm{C}}=-v_{\ell}/\sqrt{2}$. In the right section, the parameter are $\omega^{r}_{0}=\omega^{\ell}_{pol}$,  $\omega_{r}=50\,\omega^{r}_{0}$, $\Delta_{r}=\omega^{r}_{0}-\omega_{r}$, $g_{r}=g_{\ell}$, $\theta_{1}=\arctan{(2g_{r}/\Delta_{r})}$, $v_{r}=v_{\ell}/2$. Additional parameters of the numerical simulation are $\tau=10^{-3}v_{\ell}^{-1}$, $n_{\rm{max}}=2$, $\chi_{\rm max}=50$, and $L=31$.}
\label{Fig6}
\end{figure*}

Here, we describe the main contribution of our work, namely, the hybrid quantum lattice allows controlled swapping between excitation types through parameter tuning. As shown in the previous sections, depending on the regime (resonant or dispersive) and the specific conditions satisfied, the same injected excitation can propagate as a polariton, a photon, or an atomic wave. This tunability enables a dynamic excitation swap, where the type of information carrier changes during propagation.

To illustrate this mechanism, we divide the hybrid lattice into two sections, left and right, as shown in Fig.~\ref{Fig5}(a). All lattice parameters are labeled with $\ell$ and $r$ to indicate their respective sector. The excitation is injected into the left side via the activation qubit and propagates from left to right. The parameters are chosen such that the excitation propagates as one type on the left side and swaps to another type on the right side. The connection between the two regions is modeled by a coupling $\lambda_{\rm{C}}$, which acts analogously to an impedance-matching interface. Figure~\ref{Fig5}(b) shows a schematic representation of the effective energy levels, effective couplings, and the effective impedance-matching coupling. This configuration enables an excitation swap of the type polariton (left section) $\rightarrow$ atomic-wave (right section).

The activation qubit injects the excitation, which propagates as a polariton in the left section provided that the conditions detailed in Section~\ref{Sec_Polariton} are satisfied. The interface coupling $\lambda_{\rm{C}}$ is dressed by the polaritonic state on the left side, yielding the following effective impedance-matching coupling:
\begin{equation}
\lambda_{\rm{eff}}=\rho_{1-}^{\ell}\lambda_{\rm{C}}\,. 
\label{lambda_eff_L}
\end{equation}
Note that $\lambda_{\rm{C}}$ is not dressed by the right side, since the selected excitation propagating in that region is an atomic wave.
For the excitation to be fully transmitted across the interface, the reflection on the left side must be minimized to near zero. This requires that the effective hopping of the polariton match the effective impedance-matching coupling, i.e.,  $\tilde{v}_{\ell}=\lambda_{\rm{eff}}$, where $\tilde{v}_{\ell}=v_{\ell}(\rho_{1-}^{\ell})^{2}$ [see Eq.~\ref{Acop_vest_polariton}]. The last expression implies the following
\begin{equation}
    \lambda_{\rm{C}}=v_{\ell}\,\rho_{1-}^{\ell}\,.
\label{lambdaC}   
\end{equation}
Additionally, the effective impedance-matching coupling must also match the coupling between TLSs in the right section, leading to:
\begin{equation}
\lambda_{\rm{eff}}=v_{r}\,.
\label{lambda_eff_R}
\end{equation}
Substituting the Eqs.~(\ref{lambda_eff_L}) and (\ref{lambdaC}) into Eq.~(\ref{lambda_eff_R}), and using $\rho_{1-}^{\ell}=-1/\sqrt{2}$ [see Eqs.~\ref{coefficient_condition3} with $\Delta=0$], we obtain a relation between the coupling constants of nearest-neighbor TLSs in the left and right sections of the hybrid lattice, that is
\begin{equation}
v_{r}=\frac{v_{\ell}}{2}\,.
\label{vr_2}
\end{equation}

Besides, the resonance condition of the swap mechanism is given by
\begin{equation}
\omega^{r}_{0}=\omega^{\ell}_{pol}\,,
\label{Res_cond_swap}
\end{equation}
indicating that the transition frequency of the lower polaritonic branch in the left lattice side must match the transition frequency of the right-side atoms [see Fig.~\ref{Fig5}(b)].

Figure~\ref{Fig6} illustrates the dynamic excitation swap from a polariton to an atomic wave.
This figure presents the average excitation number of the $j$th TLS, $\langle\hat{\sigma}^{+}_{j}\hat{\sigma}^{-}_{j}\rangle$ (upper row), and the average photon number, $\langle \hat{n}_{j} \rangle$ (lower row), as functions of time for each $j$-th resonator unit.
Figures~\ref{Fig6}(a) and \ref{Fig6}(b) show the dynamics of these two components of the average polariton number, $\langle\hat{\mathcal{N}}_{pol,j} \rangle$, on the left side of the lattice ($2\leq j \leq 16$). In this region, both components contribute equally to the propagation of the excitation pulse. As the pulse reaches the interface and enters the right section, both photon and TLS contributions in the left region diminish nearly to zero, indicating efficient transmission with no observable reflection.
Figures~\ref{Fig6}(c) and \ref{Fig6}(d) display the corresponding dynamics in the right section ($17\leq j \leq 31$). The excitation continues to propagate as an atomic wave, as evidenced in Fig.\ref{Fig6}(c), while the photon number remains negligible, as shown in Fig.\ref{Fig6}(d). Figure~\ref{Fig6}(e) provides a complete view of the excitation dynamics across the entire system, showing the average polariton number $\langle\hat{\mathcal{N}}_{pol,j} \rangle$ as a function of time and site index. This global perspective confirms the efficient transfer of excitation from the polariton on the left to the atomic wave on the right, with negligible reflection at the interface. Additionally, the propagation velocity remains unchanged across the two regions, as evidenced by the continuous linear slope of the excitation pulse.

A sensitivity analysis of the swap mechanism is presented in Appendix~\ref{AppendixD}. 
There, we present results illustrating the swap dynamics when the matching Eq.~(\ref{lambdaC}) and resonance conditions Eq.~(\ref{Res_cond_swap}) are intentionally detuned from their optimal values.
Our results demonstrate that the swapping protocol is robust to moderate variations in the impedance-matching condition [see Fig.~\ref{FigD1}], while the resonance condition is more stringent: significant detuning leads to strong reflection and the suppression of excitation conversion [see Fig.~\ref{FigD2}].

\begin{figure}
\centering
\includegraphics[width=\columnwidth]{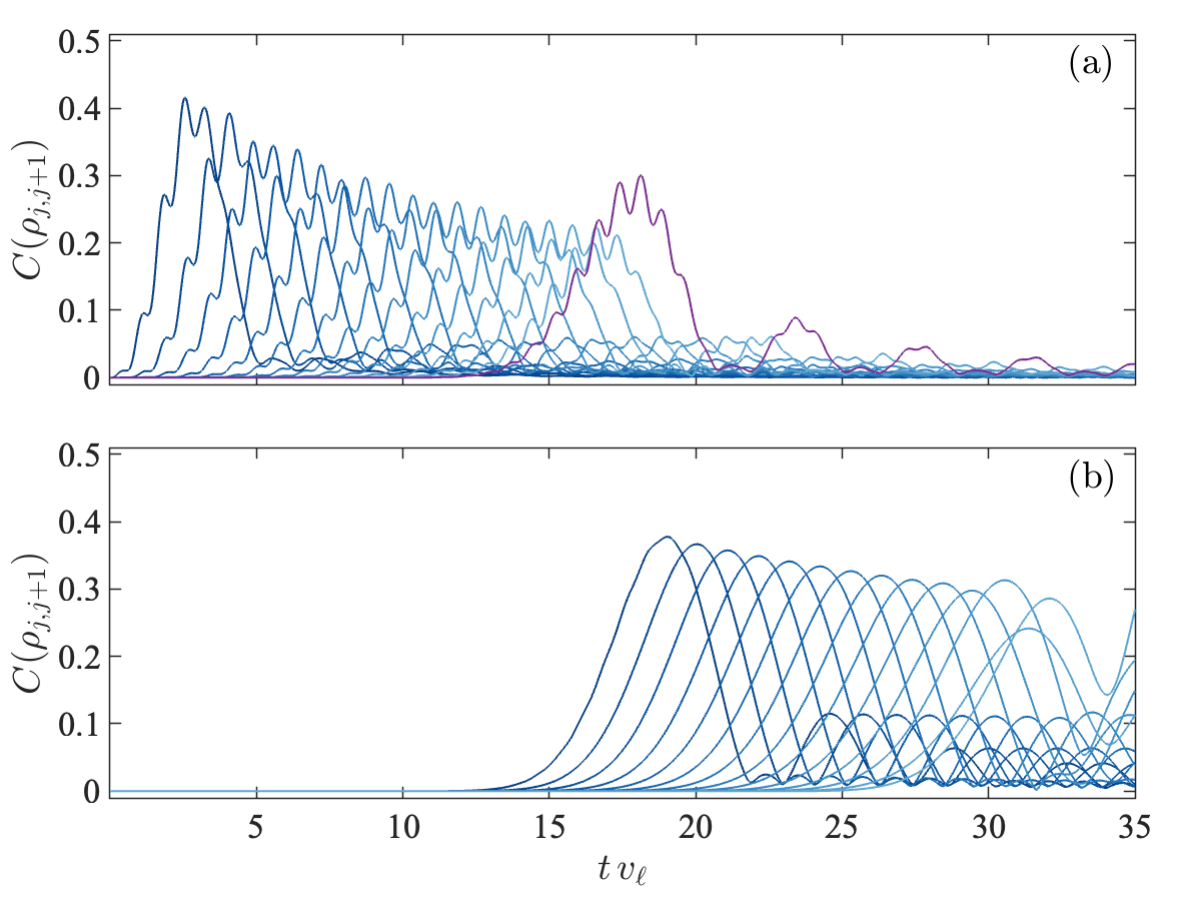}
\caption{Pairwise concurrences: $C(\rho_{j,j+1})$ as a function of time for (a) $2\leq j \leq 16$ and (b) $17\leq j \leq 30$. The purple line shows the concurrence of the impedance-matching interface between the left and right lattice sections, $C(\rho_{16,17})$. The parameters are the same as Fig.~\ref{Fig6}.}
\label{Fig7}
\end{figure}

As an additional diagnosis of the single excitation swap, we analyze entanglement via pairwise concurrences \cite{Wootters1998} between nearest-neighbor two-level systems, by tracing out photonic degrees of freedom. Given the reduced two-qubit density matrix $\rho_{j,j+1}$, the concurrence is defined as $C(\rho_{j,j+1})=\rm{max}(0,\lambda_1-\lambda_2-\lambda_3-\lambda_4)$ where $\{\lambda_k\}$ are the square roots of the eigenvalues of $R = \rho_{j,j+1}\,
      (\sigma_y \otimes \sigma_y)\,
      \rho_{j,j+1}^\ast\,
      (\sigma_y \otimes \sigma_y),
$
ordered as $\lambda_1\ge \lambda_2\ge \lambda_3\ge \lambda_4$, and $\sigma_y$ is the Pauli matrix. Figure \ref{Fig7} shows the pairwise concurrences as a function of time, where the upper panel shows concurrences of TLSs lying in the left part of the lattice. Concurrences exhibit similar patterns compared to local observables (c.f. Fig.\ref{Fig6}). This occurs due to the local interaction between the TLS and the photon. The highlighted concurrence (purple) corresponds to TLSs connecting both sides of the lattice. The lower panel shows the pairwise concurrences of TLSs lying in the right part of the lattice. Here we see an increase in the magnitude of concurrences since each TLS locally decouples from the photonic degrees of freedom, and entanglement is shared between nearest-neighbor TLSs only.

Interestingly, the excitation swap mechanism can also be realized in alternative hybrid lattice configurations. Let us consider a setup where the activation qubit is disconnected, and the system is initially prepared in a single polariton state localized at the center of the lattice ($j=(L+1)/2$ with $L$ odd). In this configuration, the lattice parameters are chosen such that the excitation propagates as a polariton to the left and as an atomic wave to the right. The two pulses propagate in opposite directions, from the center toward the lattice boundaries, while conserving the total number of excitations, a single excitation. The impedance-matching condition between the two sections, $\lambda_{\rm{C}}$ is given by Eq.~(\ref{lambdaC}), and the coupling between TLSs in the right section, $v_{r}$, is determined by Eq.~(\ref{vr_2}). 
\begin{figure}
\centering
\includegraphics[width=\columnwidth]{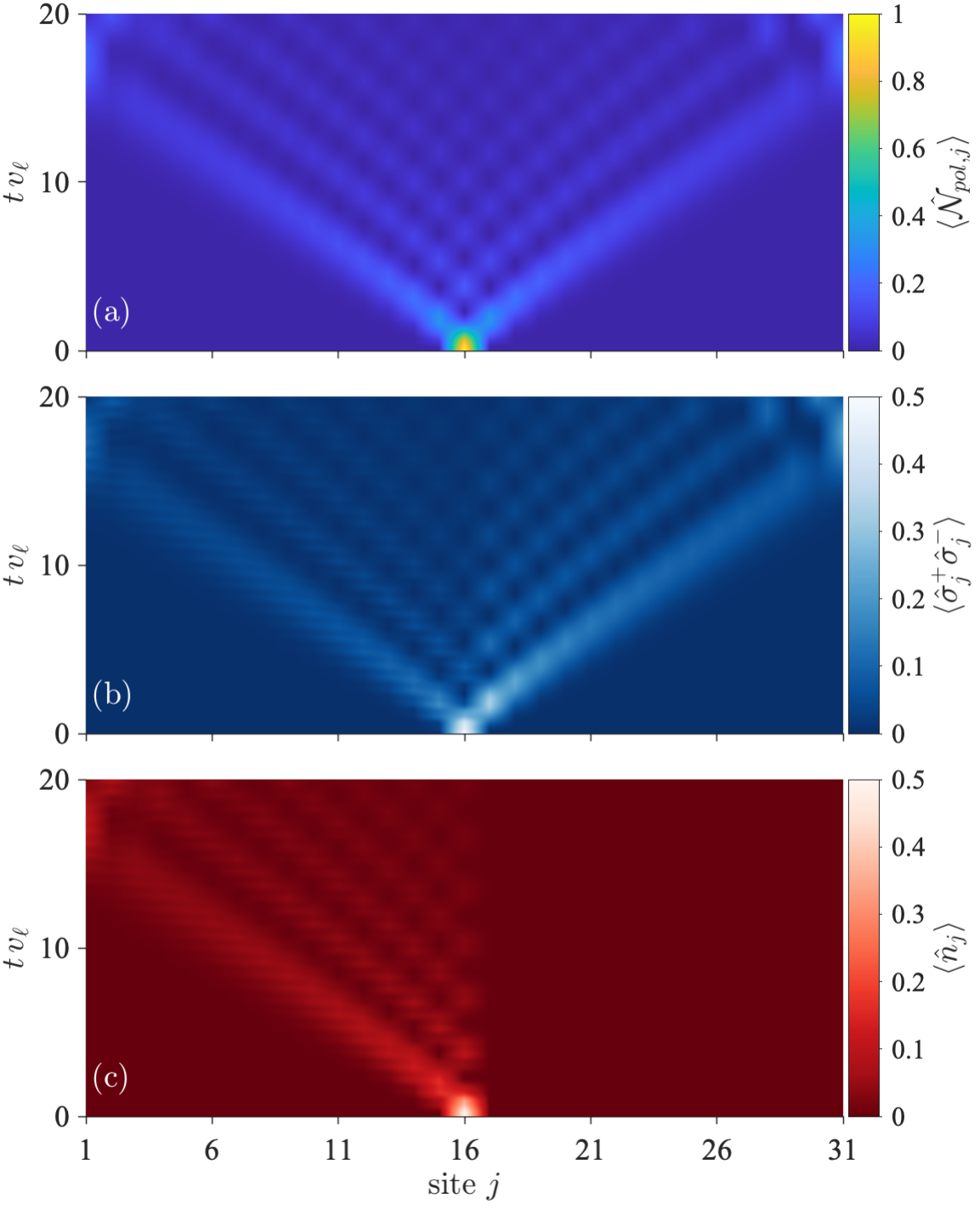}
\caption{Average polariton number, the average excitation number of the $j$th TLS and the average photon number as a function of time and the site index $j$ (with $1\leq j \leq 31$), shown in panels (a), (b), and (c), respectively.
The parameters in the left section are $\omega_{\ell}=\omega^{\ell}_{0}=1$, $\Delta_{\ell}=\omega^{\ell}_{0}-\omega_{\ell}=0$, $v_{\ell}=0.01\,\omega_{\ell}$, $g_{\ell}=4v_{\ell}$, $\omega_{\rm{A}}=\omega^{\ell}_{pol}=\omega_{\ell}-g_{\ell}$, $\lambda=\lambda_{\rm{C}}=-v_{\ell}/\sqrt{2}$. In the right section, the parameter are $\omega^{r}_{0}=\omega^{\ell}_{pol}=\omega_{\ell}-g_{\ell}$,  $\omega_{r}=50\,\omega^{r}_{0}$, $\Delta_{r}=\omega^{r}_{0}-\omega_{r}$, $g_{r}=g_{\ell}$, $\theta_{1}=\arctan{(2g_{r}/\Delta_{r})}$, $v_{r}=v_{\ell}/2$. Additional parameters of the numerical simulation are $\tau=10^{-3}v_{\ell}^{-1}$, $n_{\rm{max}}=2$, $\chi_{\rm max}=4$, and $L=31$.}
\label{Fig8}
\end{figure}

Figure~\ref{Fig8}(a) shows the average polariton number as a function of time and site index, confirming the excitation’s symmetric splitting and propagation. Figure~\ref{Fig8}(b) displays the average excitation number of the $j$th TLS, highlighting the atomic-wave character on the right half of the hybrid lattice. Figure~\ref{Fig8}(c) shows the average photon number, which remains negligible in the right section, further indicating the suppression of photonic components in the atomic-wave propagation region.

\section{Experimental Feasibility in Circuit-QED Platforms}

To assess the practical relevance of the proposed excitation-swapping mechanism, we now discuss its implementation in state-of-the-art superconducting circuit-QED platforms. In such architectures, tight-binding lattice sites can be realized using transmon qubits, while the effective tunneling between sites arises from capacitive coupling between neighboring qubits~\cite{Li2018Nov,Zha2020,Zhao2022,Song2024,CarusottoRevNat2020,Saxberg2022_Nature,Roberts_Sci2024,vrajitoarea2026}. The on-site energies can be tuned with site-resolved precision via local flux-bias lines, enabling full control over the lattice energy landscape. In addition, each transmon can be coupled to an off-resonant coplanar waveguide resonator, which mediates the light–matter interaction required to realize the modified Jaynes-Cummings-Hubbard model considered here. These capabilities suggest that the impedance-matching and resonance conditions introduced in our theoretical framework can be implemented experimentally.

In particular in Ref.~\cite{vrajitoarea2026}, the authors report the experimentally achievable parameter ranges such as: transmon tuning range frequency $\omega_{0}/2\pi \in [3,6]\,\mathrm{GHz}$, capacitive coupling between nearest-neighbor transmon sites sets a tunneling rate $J/2\pi \approx 9\,\mathrm{MHz}$, qubit-resonator dispersive couplings range $\chi_{qR}/2\pi \in[0.5, 1.6]\,\mathrm{MHz}$, qubit relaxation time $T_{1}\approx 45\,\mathrm{\mu s}$, and resonator linewidths $\kappa_{R}/2\pi\approx 100\,\mathrm{kHz}$. Using parameters within these experimentally reported ranges, the results shown in Fig.~\ref{Fig6} can be reproduced, and the excitation-swapping mechanism remains robust under realistic conditions.

In our numerical simulations, the excitation swap occurs at a characteristic dimensionless time $t_{max}=35$. Converting to physical units using the experimentally relevant coupling $J/2\pi \approx 9\,\mathrm{MHz}$, this corresponds to a timescale $T_{max}=t_{max}/J=35/2\pi\times9\,\mathrm{MHz}\approx0.6\,\mathrm{\mu s}$. This should be compared with experimentally reported qubit relaxation times of $T_{1}\approx 45\,\mathrm{\mu s}$ in circuit-QED platform~\cite{vrajitoarea2026}. Since $T_{max}\ll T_{1}$, the excitation-swapping dynamics takes place well within the coherence time of the system. This separation of timescales indicates that the proposed many-body dynamical swapping effect is, in principle, observable before decoherence significantly affects the system.

In addition, using resonator linewidths $\kappa_{R} /2\pi\approx100\,\mathrm{kHz}$ reported in Ref.~\cite{vrajitoarea2026}, the photon lifetime in the resonators is $\tau_{ph}\approx 1.6 \,\mathrm{\mu s}$. Since the characteristic timescale of the excitation swap in our simulations is $T_{max}=0.6\,\mathrm{\mu s}$, we have $T_{max}< \tau_{ph}$, indicating that the photonic component of the excitation remains coherent during the swapping process. Together with the condition $T_{max}\ll T_{1}$, this supports the feasibility of observing the proposed mechanism within current circuit-QED platforms.

\begin{figure}
\centering
\includegraphics[width=\columnwidth]{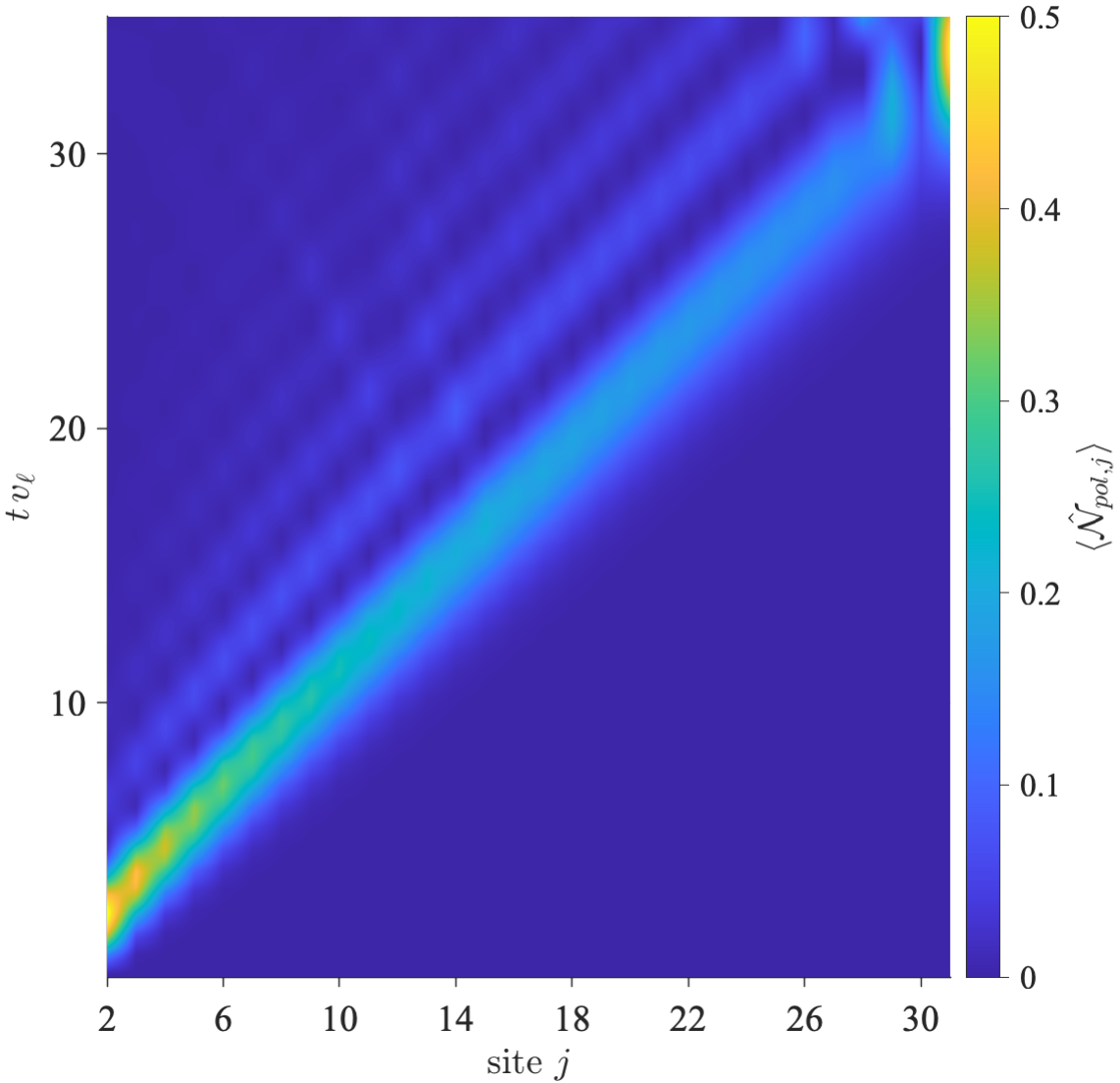}
\caption{Excitation swap using experimental parameters: Average polariton number as a function of time and site index $j$ (with $2\leq j \leq 31$).
The parameters in the left section are $\omega_{\ell}=\omega^{\ell}_{0}=1$, $\Delta_{\ell}=\omega^{\ell}_{0}-\omega_{\ell}=0$, $v_{\ell}=0.003$, $g_{\ell}=4v_{\ell}$, $\omega_{\rm{A}}=\omega^{\ell}_{pol}=\omega_{\ell}-g_{\ell}$, $\lambda=\lambda_{\rm{C}}=-v_{\ell}/\sqrt{2}$. In the right section, the parameters are $\omega^{r}_{0}=\omega^{\ell}_{pol}$, $\Delta_{r}=0.27$, $\omega_{r}=0.718$,  $g_{r}=g_{\ell}$, $\theta_{1}=\arctan{(2g_{r}/\Delta_{r})}$, $v_{r}=v_{\ell}/2$. Additional parameters of the numerical simulation are $\tau=10^{-3}v_{\ell}^{-1}$, $n_{\rm{max}}=2$, $\chi_{\rm max}=50$, and $L=31$.}
\label{Fig9}
\end{figure}

To further support the experimental feasibility, we report the numerical simulation using parameters within experimentally reported ranges. Figure~\ref{Fig9} shows the average polariton number $\langle\hat{\mathcal{N}}_{pol,j} \rangle$ as a function of time and site index. The simulations are carried out using $\omega_{0}/2\pi=3\,\mathrm{GHz}$, $J/2\pi = 9\,\mathrm{MHz}$, and $\chi_{qR}/2\pi=1.6\,\mathrm{MHz}$. All parameters in Fig.~\ref{Fig9} are normalized with respect to $\omega_{0}/2\pi$. These results show that the excitation-swapping mechanism remains robust under realistic conditions, thereby reinforcing its potential applicability in circuit-QED platforms.

\section{Conclusion}

In summary, our study demonstrates that a one-dimensional hybrid quantum lattice model, combining light-matter units with direct qubit-qubit interactions, enables controlled propagation and swapping of quantum excitations, namely, polaritons, atomic waves, and photons. By adjusting system parameters to satisfy impedance and resonance conditions, we show how these excitations can be selectively guided or coherently swapped, with polaritons serving as the interface between photonic and atomic-wave domains. For photon transport, we find that polaritonic states mediate single-excitation propagation, with negligible contribution from atomic excitations. This architecture provides a versatile platform for quantum links, where distant emitters can efficiently interact through polaritonic, atomic-wave, or photonic networks, enabling hybrid information processing and reconfigurable quantum networks. Future work should explore higher-excitation extensions, disorder effects in the hybrid lattice, and non-Markovian quantum links that communicate emitters.  

\begin{acknowledgements}
M.~A. acknowledges financial support from ANID Postdoctoral FONDECYT Grant No. 3240443. We also thank the support from Dicyt USACH under grant $042331{\rm RH\textunderscore Ayudante}$, Vicerrectoría de Investigación, Innovación y Creación, the Center for Nanoscience and Nanotechnology, CEDENNA, Project CIA250002, and Proyecto de Exploraci\'on Grant No. 13250014.

\end{acknowledgements}

\appendix
\section{Polaritonic basis}
\label{ApendixA}
The eigenstates of the Jaynes-Cummings model define the upper $(+)$ and lower $(-)$ polaritonic basis, which is used to write the Hamiltonian of our model to analyze the effective couplings to propagate the different excitations. A compact way to write the basis is to use the following notation 
\begin{equation}
    |n,\alpha\rangle_j = \gamma_{n\alpha} |\downarrow,n\rangle_j+\rho_{n\alpha}|\uparrow,n-1\rangle_j \label{pol_states}
\end{equation}
with energies 
\begin{equation}
    E^{\alpha}_n/\hbar = n\omega+\frac{\Delta}{2}\pm\frac{\Delta}{2}\sqrt{1+\frac{4g^2n}{\Delta^2}}
\end{equation}
where the polaritonic branches are indicated by the index $\alpha = \pm$, $n$ refers to the number of polaritons and the coefficients on the basis are defined as $\rho_{n+}=\cos(\frac{\theta_n}{2})$, $\gamma_{n+}=\sin(\frac{\theta_n}{2})$, $\rho_{n-}=-\gamma_{n+}$, $\rho_{n+}=\gamma_{n-}$, $\theta_n$ is defined by $\tan(\theta_n)=2g\sqrt n/\Delta$, and $\Delta = \omega_0-\omega$ is the detuning parameter. 

The ground state is identified as $|\downarrow,0\rangle = |0,-\rangle$ and the non-physical state $|0,+\rangle = |\textbf{\O}\rangle$ is a ket with all entries equal to zero. The previous determines the values of $\gamma_{0-}=1$ and $\gamma_{0+}=\rho_{0\alpha}=0$.
Also, the polaritonic creation operator can be introduced for the $j$th site as $P^{\dagger(n,\alpha)}_j = |n,\alpha\rangle_j\langle0,-|$. 

The Jaynes-Cummings model, written in the polaritonic basis reads 
\begin{equation}
    H_{\rm JC} = \sum_{j=1}^L\sum_{n=0}^\infty\sum_{\alpha=\pm}E_n^\alpha P_j^{\dagger(n,\alpha)}P_j^{(n,\alpha)}.
\end{equation}
Also, the equations for $H_A$ and $H_I$ contain terms of the raising and lowering operators. We change the representation of these terms using the completeness relation of the polaritonic basis, $\mathbf{I}_j=\sum_{n=0}^\infty\sum_{\alpha=\pm}P_j^{\dagger(n,\alpha)}P_j^{(n,\alpha)}$, dropping the site index $j$ for simplicity, we obtain
\begin{equation}
    \begin{split}
        \mathbf{I}\hspace{0.1cm}\hat{\sigma}^{+}\mathbf{I}  &= \sum_{n,m=0}^\infty \sum_{\alpha,\beta =\pm} P^{\dagger(n,\alpha)}P^{(n,\alpha)}\hat{\sigma}^{+}P^{\dagger(m,\beta)}P^{(m,\beta)}\,,\\
        &=\sum_{n,m=0}^\infty \sum_{\alpha,\beta =\pm}\langle n,\alpha|\hat{\sigma}^{+}|m,\beta\rangle P^{\dagger(n,\alpha)}P^{(m,\beta)}\,,
    \end{split} 
\end{equation}
and the matrix element is $\langle n,\alpha|\hat{\sigma}^{+}|m,\beta\rangle=\big(\gamma_{n\alpha}\langle\downarrow,n|+\rho_{n\alpha}\langle\uparrow,n-1|\big)\hat{\sigma}^{+}\big(\gamma_{m\beta}|\downarrow,m\rangle+\rho_{m\beta}|\uparrow,m-1\rangle\big)$. Considering the operator $\hat{\sigma}^{+}$ acts on the TLS and the orthogonality between the states we obtain
\begin{equation*}
    \begin{split} 
        \langle n,\alpha|\hspace{0.1cm}\hat{\sigma}^{+}\hspace{0,1cm}|m,\beta\rangle= \rho_{n\alpha}\gamma_{m\beta}\hspace{0.1cm}\delta_{n-1,m}\,.
    \end{split}
\end{equation*}
Here, the Kronecker delta is chosen such that $m=n-1$
\begin{equation*}
    \mathbf{I}\hspace{0.1cm}\hat{\sigma}^{+}\mathbf{I} = \sum_{n=1}^\infty \sum_{\alpha,\beta =\pm}\rho_{n\alpha}\gamma_{n-1\beta} P^{\dagger(n,\alpha)}P^{(n-1,\beta)}\,,
\end{equation*}
and taking into account $\hat{\sigma}^{-} = (\hat{\sigma}^{+})^\dagger$, we have 
\begin{equation*}
    \begin{split}
        \hat{\sigma}_{j}^{+}\hat{\sigma}_{j+1}^{-} = \sum_{n,m=1}^{\infty}\sum_{\substack{\alpha,\alpha'=\pm\\\beta,\beta'=\pm}}\big(&\rho_{n\alpha}\gamma_{n-1\beta}\rho_{m\alpha'}\gamma_{m-1\beta'}
         P_j^{\dagger(n,\alpha)}\\&P_j^{(n-1,\beta)}P_{j+1}^{\dagger(m-1,\beta')}P_{j+1}^{(m,\alpha')}\big)\,,
    \end{split}
\end{equation*}
and accordingly, their complex conjugate. Considering the previous, with abbreviated notation $t_{n}^{\alpha\beta}=\rho_{n\alpha}\gamma_{n-1\beta}$, the $H_I$ Hamiltonian on the polaritonic basis is
\begin{equation}
\begin{split}
    H_I =\hbar v\sum_{j}^{L-1}\sum_{n,m=1}^{\infty}\sum_{\substack{\alpha,\alpha'=\pm\\\beta,\beta'=\pm}}\big[t_{n}^{\alpha\beta}t_{m}^{\alpha'\beta'}&P_{j}^{\dagger(n,\alpha)}P_{j}^{(n-1,\beta)}\\P_{j+1}^{\dagger(m-1,\beta')}P_{j+1}^{(m,\alpha')}+ {\rm h.c} \big]\,.
\end{split}
\end{equation}
Now we move to the interaction picture. We compute $\mathcal{H_I} = U^{\dagger}VU$ with the evolution operator $U(t) = \exp(-\frac{i}{\hbar}\sum_{j}^{L}H_{{\rm JC}_j}t)$. Since the operators of different sites commute, we have local evolution operators for each site as $U(t)=\prod_{j}^{L}U_{j}(t)$ with 
\begin{equation*}
\begin{split}
U_{j}(t)&=\exp\bigg({-\frac{i}{\hbar}\sum_{n=1}^{\infty}\sum_{\alpha=\pm}F_{n}^{\alpha}(t)P_{j}^{\dagger(n,\alpha)}P_{j}^{(n,\alpha)}}\bigg)\,,\\
&= {\sum_{n=1}^{\infty}\sum_{\alpha=\pm}\mathrm{e}^{-\frac{i}{\hbar}F_{n}^{\alpha}(t)}P_{j}^{\dagger(n,\alpha)}P_{j}^{(n,\alpha)}}\,,
\end{split}
\end{equation*}
where $F_{n}^{\alpha}(t)=E_{n}^{\alpha}t$. As a result of the previous, we only need to compute two terms: $\textrm{I})\hspace{0.1cm}U_j^{\dagger}(t)P_{j}^{\dagger(n,\alpha)}P_{j}^{(n-1,\beta)}U_{j}(t)$ and $\textrm{II)}\hspace{0.1cm}U_{j+1}^{\dagger}(t)P_{j+1}^{\dagger(m-1,\beta')}P_{j+1}^{(m,\alpha')}U_{j+1}(t)$. Without loss of generality, we calculate the terms $\textrm{I}$ and $\textrm{II}$ for a general index site $j=l$, and we highlight that the polaritonic branch indexes and the number of polaritons are independent for every operator of these terms. Consequently, for the local evolution operators, we introduce the indexes $\eta$ and $\mu$ for the number of excitations and $\zeta$ and $\xi$ for the polaritonic branches.

\begin{equation*}
\begin{split} 
    \textrm{I})\hspace{0.1cm}U^{\dagger}&P^{\dagger(n,\alpha)}P^{(n-1,\beta)}U\\ =&\sum_{\eta,\mu=1}^{\infty}\sum_{\zeta,\xi=\pm}\mathrm{e}^{\frac{i}{\hbar}F_{\eta}^{\zeta}}\mathrm{e}^{-\frac{i}{\hbar}F_{\mu}^{\xi}}P^{\dagger(\eta,\zeta)}P^{(\eta,\zeta)}\\
    &P^{\dagger(n,\alpha)}P^{(n-1,\beta)}P^{\dagger(\mu,\xi)}P^{(\mu,\xi)} \,,\\
    =& \sum_{\eta,\mu=1}^{\infty}\sum_{\zeta,\xi=\pm}\mathrm{e}^{\frac{i}{\hbar}(F_{\eta}^{\zeta}-F_{\mu}^{\xi})}\delta_{\eta,n}\delta_{\zeta,\alpha}\delta_{n-1,\mu}\delta_{\beta,\xi}\ket{\eta,\zeta}\bra{\mu,\xi}\,.
\end{split}
\end{equation*}
The last line of the equation above is obtained by considering the explicit form of the polaritonic operators in terms of the states and the orthogonality of the basis. Similarly, the second term reads
\begin{equation*}
    \begin{split}
        \textrm{II)}\hspace{0.1cm}U^{\dagger}&P^{\dagger(m-1,\beta')}P^{(m,\alpha')}U\\
        =&\sum_{\eta',\mu'=1}^{\infty}\sum_{\zeta',\xi'=\pm}\mathrm{e}^{\frac{i}{\hbar}(F_{\eta'}^{\zeta'}-F_{\mu'}^{\xi'})}\\&\delta_{\eta',m-1}\delta_{\zeta',\beta'}\delta_{m,\mu'}\delta_{\alpha',\xi'}\ket{\eta',\zeta'}\bra{\mu',\xi'}\,.
    \end{split}
\end{equation*}
Solving the deltas such that all is in terms of $n,\hspace{0.1cm}m,\hspace{0.1cm}\alpha\hspace{0.1cm}(\alpha')$ and $\beta \hspace{0.1cm}(\beta')$ and transforming the states back into the polaritonic operator form, we obtain $\mathcal{H_I}=U^{\dagger}VU$. For the sake of notation space, we define $\mathcal{T}^{\alpha\beta\alpha^{'}\beta^{'}}_{nm}= vt_{n}^{\alpha\beta}t_{m}^{\alpha'\beta'}$
\begin{equation}
    \begin{split}
        \mathcal{H_I} &= \hbar\sum_{j=1}^{L-1}\bigg[\sum_{\substack{n,\\m}=1}^{\infty}\sum_{\substack{\alpha,\alpha'=\pm\\\beta,\beta'=\pm}}\mathcal{T}^{\alpha\beta\alpha^{'}\beta^{'}}_{nm}\mathrm{e}^{\frac{i}{\hbar}(F_{n}^{\alpha}-F_{n-1}^{\beta}+F_{m-1}^{\beta'}-F_{m}^{\alpha'})}\\&P_j^{\dagger(n,\alpha)}P_j^{(n-1,\beta)}P_{j+1}^{\dagger(m-1,\beta')}P_{j+1}^{(m,\alpha')} +  {\rm h.c.} \bigg]\,.
    \end{split}\label{HamInt_ApA}
\end{equation}
We choose to propagate polaritons through the lower branch; therefore, propagation through the upper polaritonic branch must be suppressed. Considering two general contiguous sites of the chain in which there is a polariton, $\ldots\ket{1,-}_{l}\ket{0,-}_{l+1}\ldots$, we expect that the propagation of the polariton will be $\ldots\ket{0,-}_{l}\ket{1,-}_{l+1}\ldots$ instead of $\ldots\ket{0,-}_{l}\ket{1,+}_{l+1}\ldots$. The hermitian conjugate part in the Hamiltonian of equation \eqref{HamInt_ApA} leads to the term $\ket{0,-}_{l}\ket{1,\alpha'}_{l+1}$. The indices for propagation through the upper branch correspond to $n=m=1$, $\alpha=\beta'=\beta=-$, and $\alpha'=+$. 
The latter implies two things. In the first place, for the energy values in the exponential term of equation \eqref{HamInt_ApA}, we have $\exp{\big(-\frac{i}{\hbar}[E_1^{-}-E_1^{+}]t\big)}=\exp{\big(\frac{2i}{\hbar}\big[\sqrt{\Delta^2/4+g^2}}\big]t\big)$. In the resonant regime, we obtain $\exp{\big(\frac{2i}{\hbar}g t\big)}$. In second place $\mathcal{T}^{--+-}_{11}= vt_{1}^{--}t_{1}^{+-}=v(\rho_{1-}\gamma_{0-}\rho_{1+}\gamma_{0-})=-\frac{v}{2}\sin{\theta_1}$. Remembering the definition $\tan{\frac{2g\sqrt{n}}{\Delta}}=\frac{\sin{\theta_n}}{\cos{\theta_n}}$, and since $\Delta=0$, then $\tan{\theta_{1}\xrightarrow{}\infty}$. For this reason, we choose $\theta_1=(2k+1)\pi/2;k\in\mathbb{Z}$. Due to the rotating-wave approximation, we obtain the condition to suppress the fast oscillating terms as $2g \gg\mathcal{T}^{--+-}_{11}$, and suppress the propagation of polaritons through the upper branch. In terms of the hopping strength amplitude $|v|$, the condition for suppressing the upper polariton branch reads
\begin{equation}
    g\gg\frac{|v|}{4}\,.
\label{gcondition}    
\end{equation}

\renewcommand{\thefigure}{B\arabic{figure}}
\setcounter{figure}{0}
\begin{figure}
\centering
\includegraphics[width=\columnwidth]{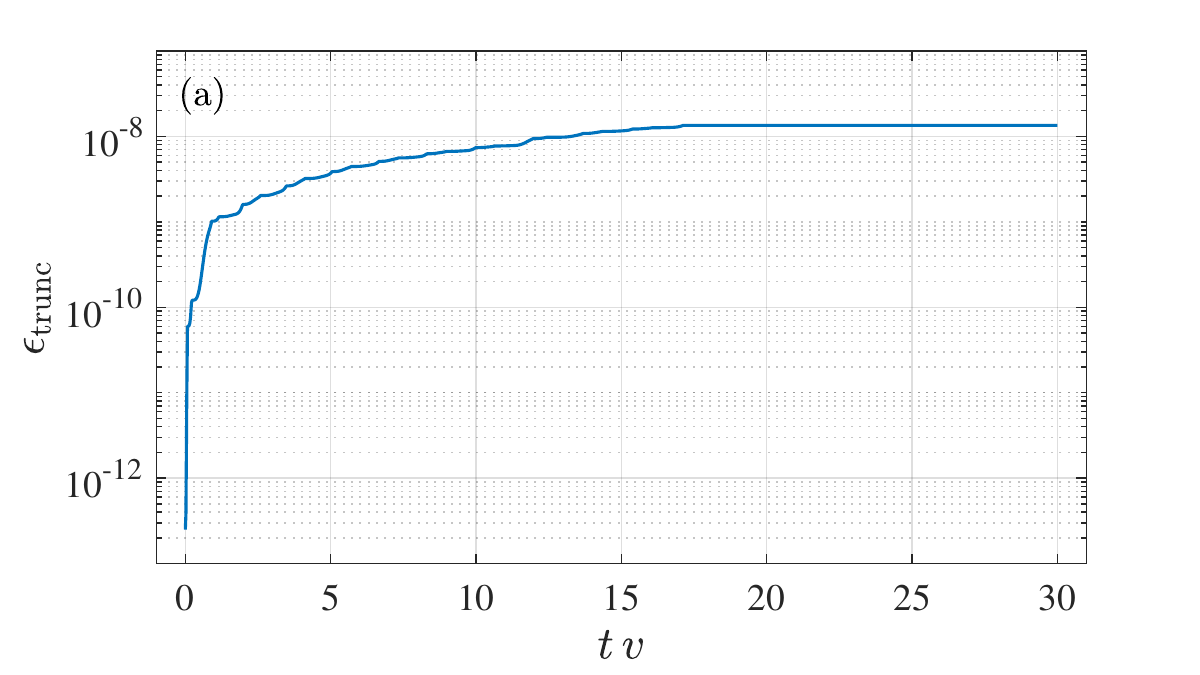}
\includegraphics[width=\columnwidth]{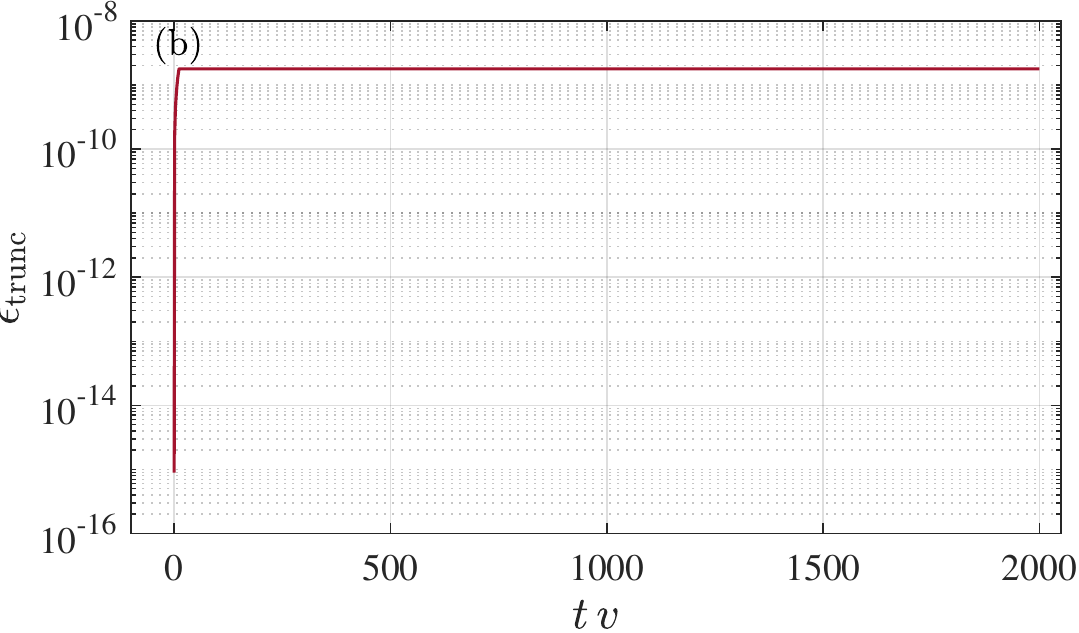}
\includegraphics[width=\columnwidth]{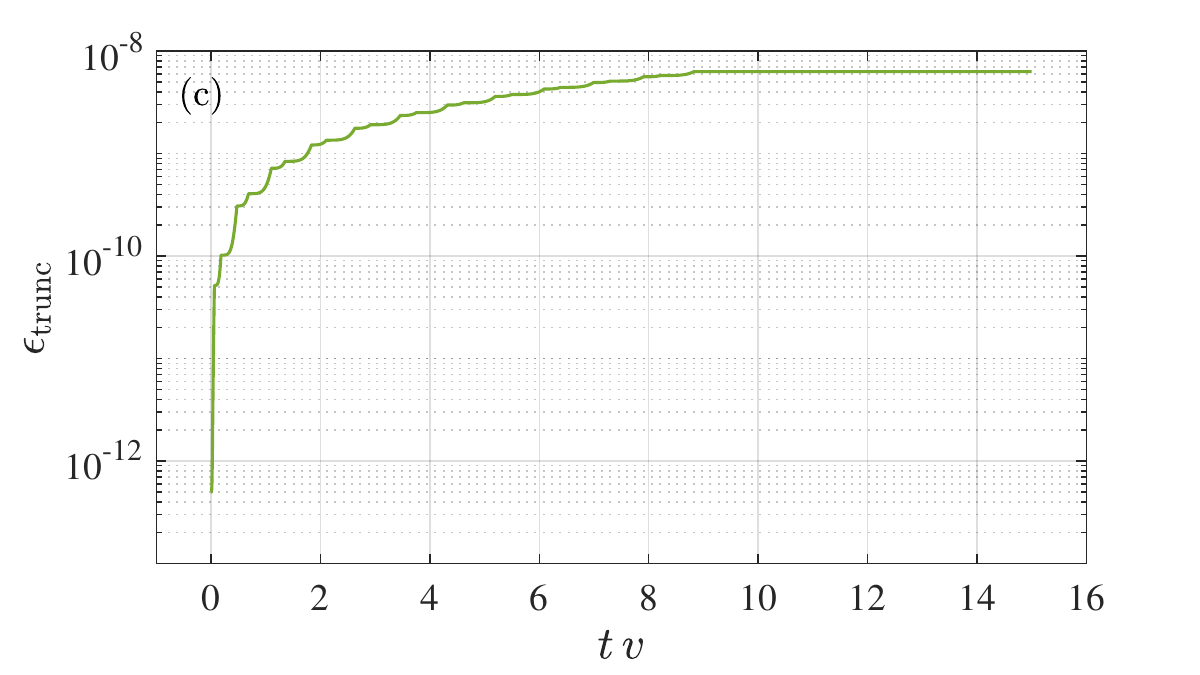}
\caption{TEBD-truncation error as a function of time for (a) polariton, (b) photon, and (c) spin-wave propagation. The vertical axis scale is logarithmic using base 10. TEBD-truncation errors correspond to the cases of Fig.~\ref{Fig2}, Fig.~\ref{Fig3}, and Fig.~\ref{Fig4} in the main manuscript, respectively. 
}
\label{FigB}
\end{figure}

\section{Truncation error}
\label{AppendixB}

In the TEBD algorithm, a two-site wave function's split and truncate procedure in a mixed canonical form involves the singular value decomposition (SDV) of a 4-rank multidimensional array. The SVD allows us to compute ${\rm norm_{old}}=\sqrt{\sum_a\lambda_a^2}$ where $\lambda_a$ are the Schmidt coefficients. Then, we proceed with the truncation by keeping the Schmidt coefficients larger than a given threshold $\epsilon_0$, which we fix to $\epsilon_0 = 10^{-6}$ in our algorithm. The new bond dimension of the updated matrix product state (MPS) is computed as ${\rm min}(\chi_{\rm max},N_{\rm kept})$, where $N_{\rm kept}$ is the number of kept singular values, and $\chi_{\max}$ is the maximum bond dimension of the MPS, which we fix to $\chi_{\rm max}=4$ for single excitation propagation and $\chi_{\rm max}=50$ for the excitation swap. Introducing the threshold value $\epsilon_0$ allows an adaptive bond dimension along the quantum lattice, thus saving computational time. Once the truncation has been done, we compute ${\rm norm_{new}}=\sqrt{\sum_{a=1}^{N_{\rm kept}}\lambda_a^2}$ and $1-2\epsilon_n$, where $\epsilon_n=1-({\rm norm_{new}}/{\rm norm_{old}})^2$. The described procedure is done each time a two-body unitary is applied following the Suzuki-Trotter decomposition \cite{hatano2005} of the evolution operator, and the quantity $1-2\epsilon_n$ is saved in a vector. The total truncation error is computed as $\epsilon_{\rm trunc}=1-\prod_n(1-2\epsilon_n)$ \cite{tenpy2024}.   

Figures~\ref{FigB}(a),~\ref{FigB}(b), and ~\ref{FigB}(c) show the truncation error as a function of time of polariton, photon, and spinwave propagation, respectively. Each plot shows a negligible truncation error $\epsilon_{\rm trunc}\lesssim 10^{-8}$, a direct consequence of a single excitation propagating along the lattice. 

\section{Polaritonic branch populations}
\label{AppendixC}
\renewcommand{\thefigure}{C\arabic{figure}}
\setcounter{figure}{0}
\begin{figure}
\centering
\includegraphics[width=\columnwidth]{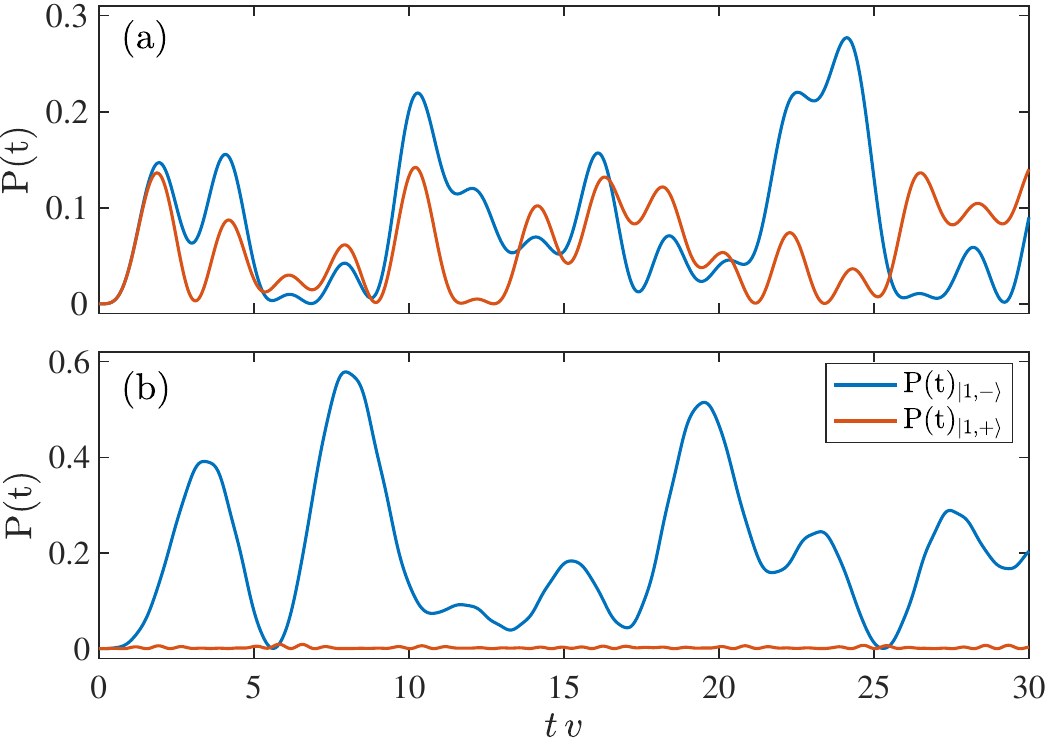}
\caption{Polaritonic branch populations as a function of time in units of $v$, (a) for $g = v/4$ and (b) $g = 4v$, satisfying the condition of Eq.~(\ref{gcondition}). The numerical simulation parameters are detailed in Fig.~\ref{Fig2} of the main manuscript.}
\label{FigC1}
\end{figure}

Here, we present numerical results for the population of the upper and lower polaritonic branches. We are interested in verifying the condition $g\gg \lvert v\lvert/4$ where the upper polaritonic branch is suppressed and the lower polaritonic branch is populated by activation qubit frequency condition $\omega_A = \omega - g$. We numerically solve the time-dependent Schrödinger equation using exact diagonalization. We thus consider a reduced system of $L=4$ sites, restricting the photonic Hilbert space to at most $n_{\rm max}=2$ photons per resonator. This choice is justified by our observation that only the $n = 1$ subspace is appreciably populated, with higher Fock states contributing negligibly to the total. 

In figure~\ref{FigC1}, we show the polaritonic branch populations for two cases: when $g = v/4$ and $g = 4v$, for the resonator in the site $j = 3$. In the first case, we observe that the population of the upper polaritonic branch is comparable to that of the lower branch, reaching approximately 10\%. This suggests that the polaritonic excitation propagates through both branches simultaneously. In the second case, only the lower polaritonic branch is significantly populated, thus confirming the regime $g \gg \frac{|v|}{4}$, where the dynamics is dominated by a single polaritonic branch.

\section{Swap sensitivity}
\label{AppendixD}
\renewcommand{\thefigure}{D\arabic{figure}}
\setcounter{figure}{0}

\begin{figure*}[t]
\centering
\includegraphics[width=\textwidth]{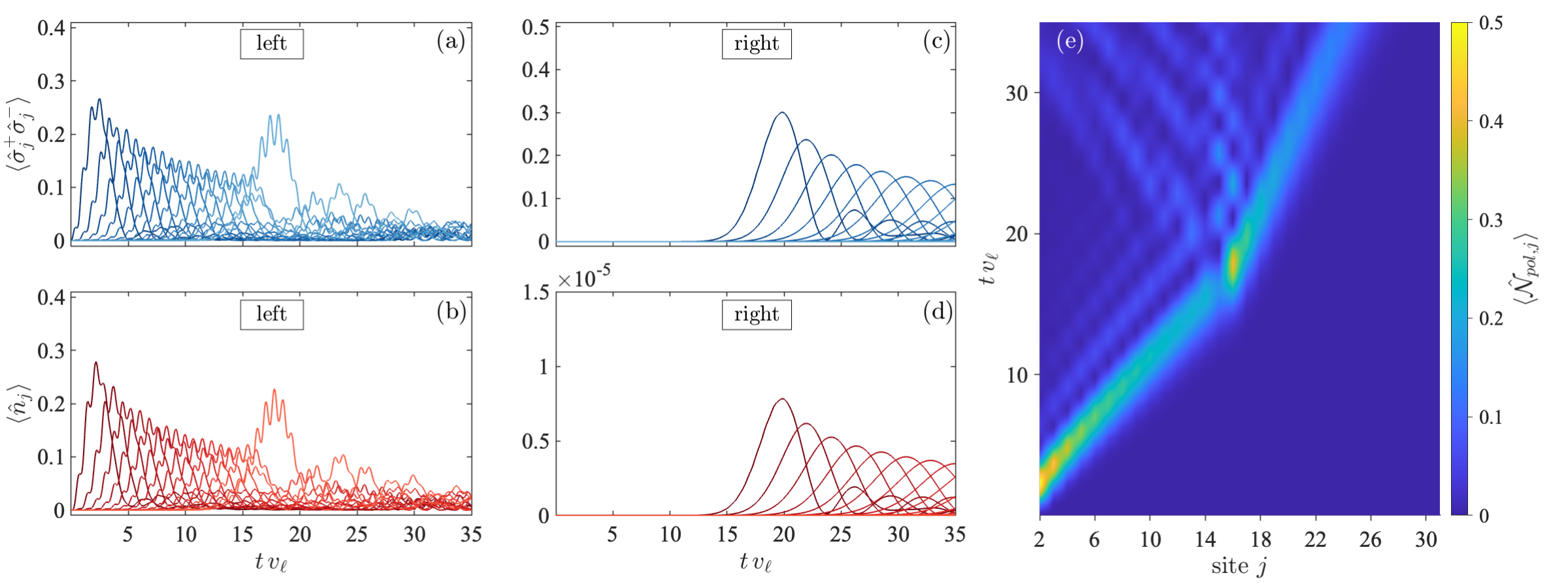}
\caption{Swap parameters sensitivity, polariton $\rightarrow$ atomic-wave: Average excitation number of the $j$th TLS and the average photon number as functions of time for the left section ($\ell$), shown in panels (a) and (b), and for the right section ($r$), shown in panels (c) and (d), respectively. (e) Average polariton number as a function of time and the site index $j$ (with $2\leq j \leq 31$).
The parameters in the left section are $\omega_{\ell}=\omega^{\ell}_{0}=1$, $\Delta_{\ell}=\omega^{\ell}_{0}-\omega_{\ell}=0$, $v_{\ell}=0.05\,\omega_{\ell}$, $g_{\ell}=4v_{\ell}$, $\omega_{\rm{A}}=\omega^{\ell}_{pol}=\omega_{\ell}-g_{\ell}$, $\lambda=-v_{\ell}/\sqrt{2}$, $\lambda^{\prime}_{\rm{C}}=-\,v_{\ell}/2\sqrt{2}$. In the right section, the parameter are $\omega^{r}_{0}=\omega^{\ell}_{pol}$,  $\omega_{r}=50\,\omega^{r}_{0}$, $\Delta_{r}=\omega^{r}_{0}-\omega_{r}$, $g_{r}=g_{\ell}$, $\theta_{1}=\arctan{(2g_{r}/\Delta_{r})}$, $v_{r}=v_{\ell}/2$. Additional parameters of the numerical simulation are $\tau=10^{-3}v_{\ell}^{-1}$, $n_{\rm{max}}=2$, $\chi_{\rm max}=50$, and $L=31$.}
\label{FigD1}
\end{figure*}

We present the swap dynamics when the matching and resonance conditions are detuned from their optimal values [see Eqs.(\ref{lambdaC}) and~(\ref{Res_cond_swap}) in the main text]. Figure~\ref{FigD1} shows the excitation dynamics when the impedance-matching condition is modified to $\lambda^{\prime}_{\rm{C}}=\lambda_{\rm{C}}/2$, while all remaining parameters are identical to those used in Fig.~\ref{Fig6}. In this case, the conversion from a polariton to an atomic wave still occurs, but the process is no longer ideal: a partial reflection appears on the left side of the interface. Moreover, in the right section of the lattice, the atomic wave propagates with a reduced group velocity, as indicated by the smaller slope of the wavefront in Fig.~\ref{FigD1}(e).

Figure~\ref{FigD2} highlights the sensitivity of the swapping mechanism to the resonance condition, which in the optimal case requires the polariton frequency on the left side to match the transition frequency of the TLSs on the right, i.e., $\omega^{r}_{0}=\omega^{\ell}_{pol}$. In Fig.~\ref{FigD2}, we detune this condition by setting $\omega^{r}_{0}=\omega^{\ell}_{pol}/2$, while keeping all other parameters equal to those of Fig.~\ref{Fig6}. Under this detuned resonance, the incoming polariton wave is completely reflected, and the atomic excitation fails to propagate into the right-hand lattice. This behavior is consistent with Figs.~\ref{FigD2}(c)–\ref{FigD2}(e), which show the absence of excitation transfer across the interface.

\renewcommand{\thefigure}{D\arabic{figure}}
\setcounter{figure}{1}

\begin{figure*}[t]
\centering
\includegraphics[width=\textwidth]{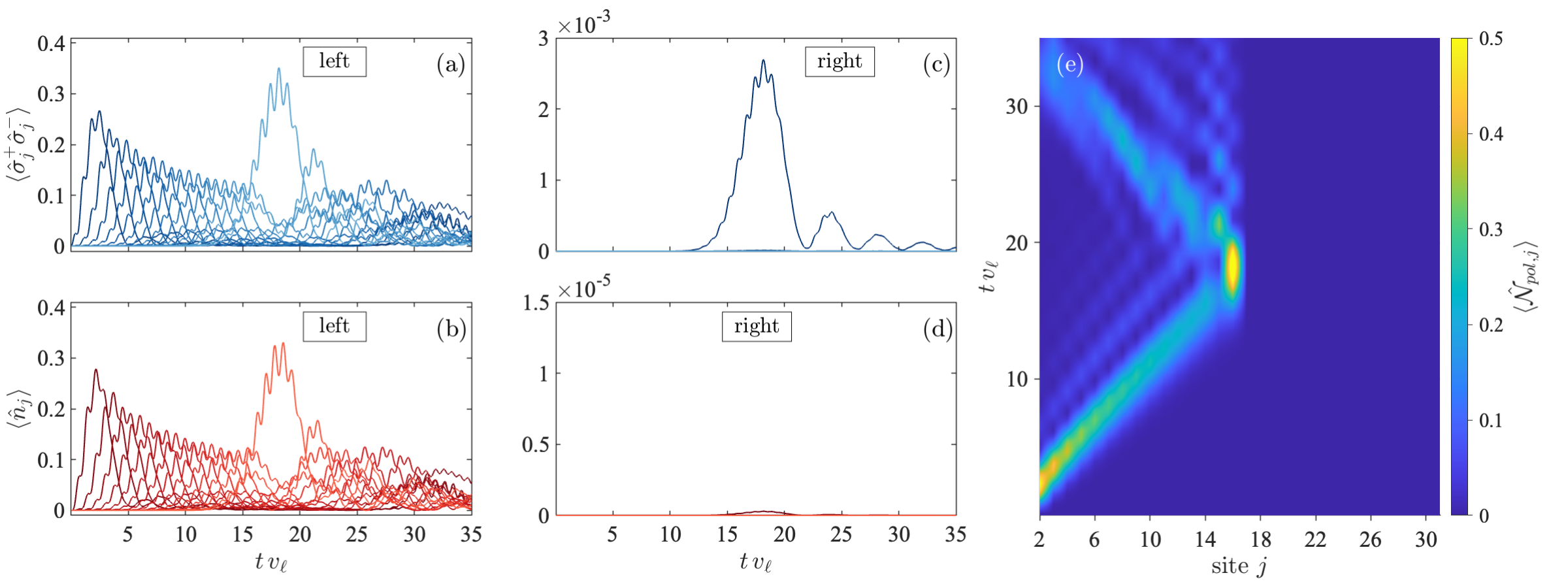}
\caption{Swap parameters sensitivity, polariton $\rightarrow$ atomic-wave: Average excitation number of the $j$th TLS and the average photon number as functions of time for the left section ($\ell$), shown in panels (a) and (b), and for the right section ($r$), shown in panels (c) and (d), respectively. (e) Average polariton number as a function of time and the site index $j$ (with $2\leq j \leq 31$).
The parameters in the left section are $\omega_{\ell}=\omega^{\ell}_{0}=1$, $\Delta_{\ell}=\omega^{\ell}_{0}-\omega_{\ell}=0$, $v_{\ell}=0.05\,\omega_{\ell}$, $g_{\ell}=4v_{\ell}$, $\omega_{\rm{A}}=\omega^{\ell}_{pol}=\omega_{\ell}-g_{\ell}$, $\lambda=\lambda_{\rm{C}}=-v_{\ell}/\sqrt{2}$. In the right section, the parameter are $\omega^{r}_{0}=\omega^{\ell}_{pol}/2$, $\omega_{r}=50\,\omega^{r}_{0}$, $\Delta_{r}=\omega^{r}_{0}-\omega_{r}$, $g_{r}=g_{\ell}$, $\theta_{1}=\arctan{(2g_{r}/\Delta_{r})}$, $v_{r}=v_{\ell}/2$. Additional parameters of the numerical simulation are $\tau=10^{-3}v_{\ell}^{-1}$, $n_{\rm{max}}=2$, $\chi_{\rm max}=50$, and $L=31$.}
\label{FigD2}
\end{figure*}
\vspace{1.5cm}
\bibliography{references}

\end{document}